# Readability Research: An Interdisciplinary Approach

Sofie Beier[1]
Sam Berlow[2]
Esat Boucaud[3]
Zoya Bylinskii[4]
Tianyuan Cai[4]
Jenae Cohn[5]
Kathy Crowley[6]
Stephanie L. Day[3]
Tilman Dingler[7]
Jonathan Dobres[8]
Jennifer Healey[4]
Rajiv Jain[4]
Marjorie Jordan[6]
Bernard Kerr[4]
Qisheng Li[9]
Dave B. Miller[3]
Susanne Nobles[10]
Alexandra Papoutsaki[11]
Jing Qian[12]
Tina Rezvanian[4]
Shelley Rodrigo[13]
Ben D. Sawyer[3]
Shannon M. Sheppard[14]
Bram Stein[4]
Rick Treitman[4]
Jen Vanek[15]
Shaun Wallace[12]
Benjamin Wolfe[16]

*Authors presented in alphabetical author for this collaborative work*

[1]Centre for Visibility Design, Royal Danish Academy, Copenhagen, Denmark
[2]Typography for Good
[3]Virtual Readability Lab, University of Central Florida, Orlando, FL
[4]Adobe Inc., San Francisco, CA
[5]California State University, Sacramento, Sacramento, CA
[6]Readability Matters
[7]University of Melbourne, Melbourne, Australia
[8]Sonos, Inc., Boston, MA
[9]Paul G. Allen School of Computer Science & Engineering, University of Washington
[10]ReadWorks, Brooklyn, NY

[11]Department of Computer Science, Pomona College Claremont, CA
[12]Brown University, Providence, RI
[13]University of Arizona, Tucson, AZ
[14]Chapman University, Department of Communication Sciences & Disorders, Orange, CA
[15]World Education, Digital Learning and Research, Boston, MA
[16]University of Toronto Mississauga, Mississauga ON, Canada

Corresponding Author: Ben D. Sawyer, sawyer@ucf.edu, Department of Industrial Engineering and Management Systems, 4000 Central Florida Blvd.,P.O. BOX 162993, Orlando, FL 32816-2993

# Abstract

Readability is on the cusp of a revolution. Fixed text is becoming fluid as a proliferation of digital reading devices rewrite what a document can do. As past constraints make way for more flexible opportunities, there is great need to understand how reading formats can be tuned to the situation and the individual. We aim to provide a firm foundation for readability research, a comprehensive framework for modern, multi-disciplinary readability research. Readability refers to aspects of visual information design which impact information flow from the page to the reader. Readability can be enhanced by changes to the set of typographical characteristics of a text. These aspects can be modified on-demand, instantly improving the ease with which a reader can process and derive meaning from text. We call on a multi-disciplinary research community to take up these challenges to elevate reading outcomes and provide the tools to do so effectively.

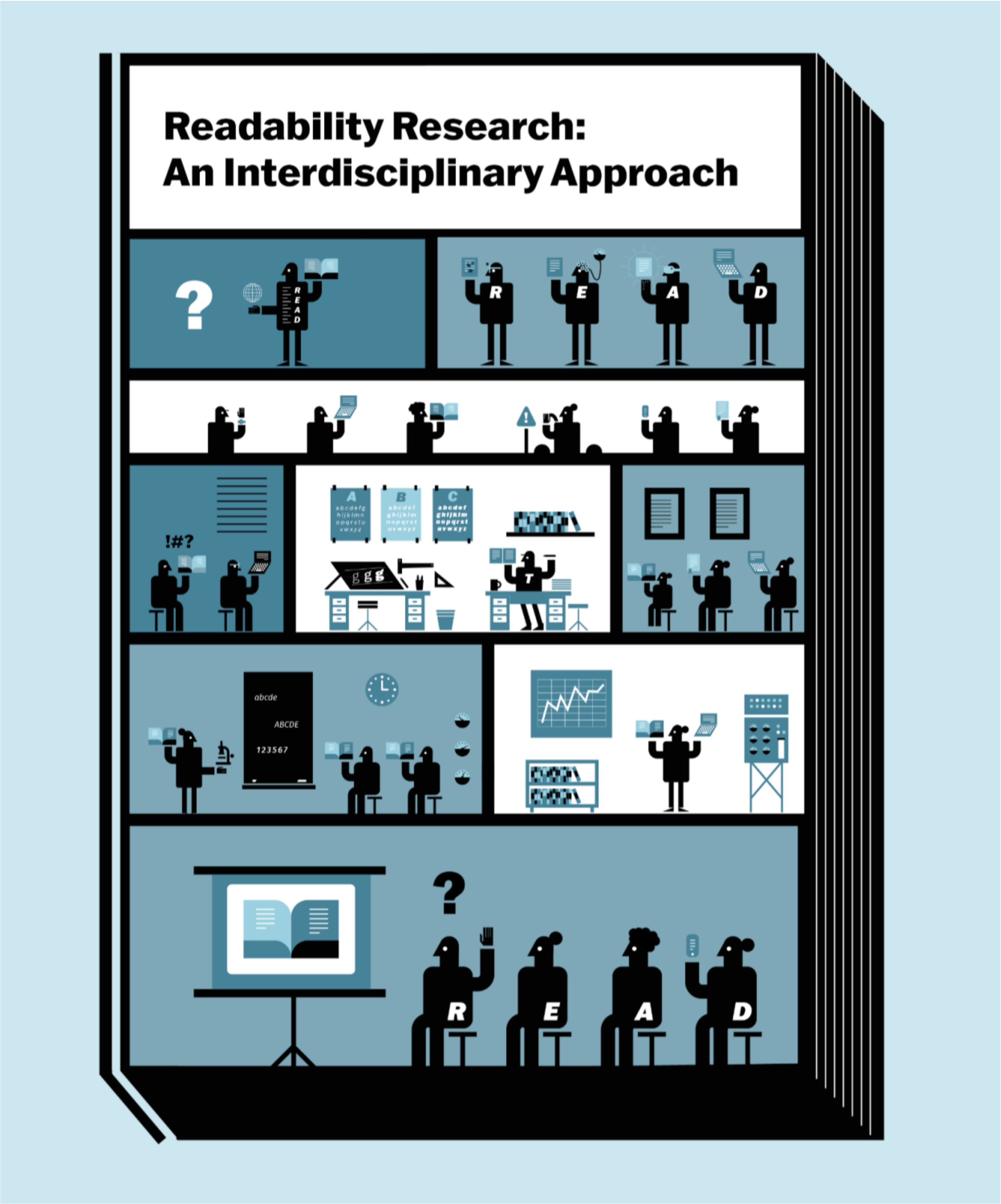

**Figure 1.** Readability is aspects of visual information design which impact information flow from the page to the reader. It can be enhanced, for example by targeted changes to typographical characteristics, or to content. A growing group of researchers are working to understand how and why these changes help.

# 1. Introduction

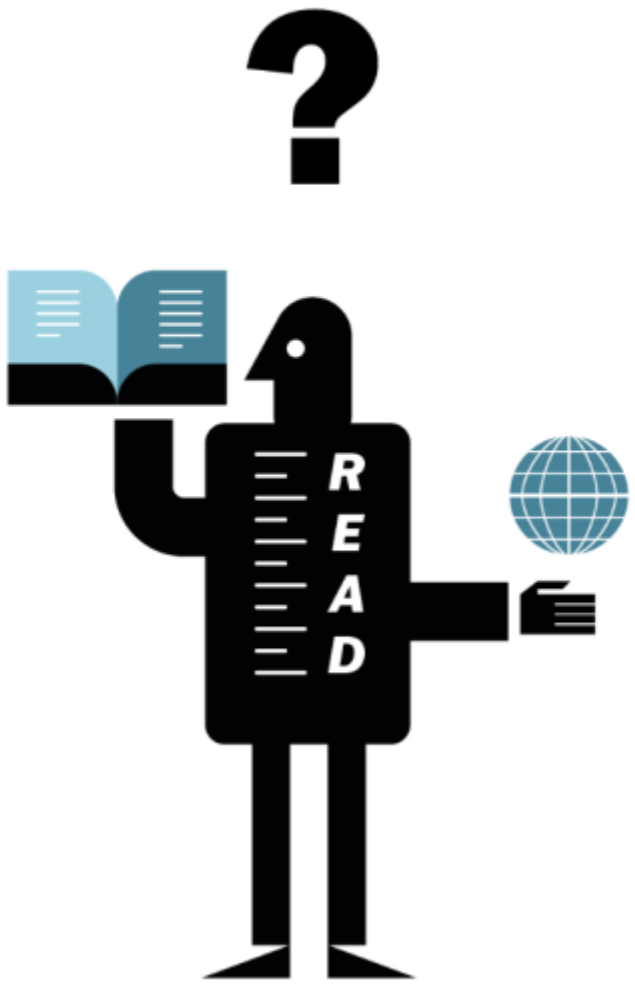

**Figure 2.** Digital technologies allow dynamic, adaptive reader control of font size, font choice, screen polarity, spacing, and content itself, aspects traditionally controlled by the publisher. New digital technology enables such changes on-demand, instantly enhancing individual readability.

Readability (Fig. 1) is a growing focus because reading has fundamentally changed (Fig. 2). The amount of reading content available today is more than has ever existed, and it grows at an ever-increasing rate. The pressure to ingest vast quantities of information quickly has never been greater. Additionally, reading content is distributed across highly variable publishing platforms, which has changed what and how we read. The wider commercialization of smartphones, e-books, and lightweight laptops over the last 15 years have diversified where and when we read. Now more than ever, improving the modern reading experience has enormous practical and societal significance.

Modern reading is digital, and this has introduced a fundamental paradigm shift. To date, authors, publishers, and designers have been in control of the reading experience. This is no longer the case as evidenced by the multitude of device types, screen qualities, and software settings available to the reader. Depending on the technologies we use, we - the readers - can gain control over font size, screen polarity, spacing, font choice, and other formatting choices. Amazon’s Kindle, Apple’s iBooks, Microsoft’s Immersive Reader, Adobe’s Liquid Mode, and modern web browsers all provide some of these controls. Because our devices are always with us, we now grab bits of reading here and there (Wolf, 2018): some of us check social media, others their news feeds, and still others their ebooks, news apps, or PDFs for class or the next meeting. Increasingly, we read in interludes, and with frequent interruptions, a considerably different form of reading than sitting down at length with a novel or newspaper. Existing research on reading has concentrated on fixed-format pages - either books or electronic facsimiles of books like PDFs. Recent studies indicate that we may be on the cusp of dramatically changing the way people read - making it much easier for struggling readers to read and for good readers to read even more efficiently by changing and, more significantly, personalizing the readability of the text. Indeed, the authors have seen children, both good and struggling readers, become more fluent with customized text formats (Crowley & Jordan, 2019a; Crowley & Jordan, 2019b) and adult readers add as much as 10 pages an hour when optimizing digital text for personalized reading (Wallace et al., 2020a; Wallace et al., 2020b).

***Readability*** encapsulates those properties of a document that determine the ease and success with which individual readers decipher, process, and make meaning of the text read. While this may involve aspects of the document's content, structure, or layout more generally, our focus in this paper is on the typographical features of the text, which includes font choice, size, spacing, and related attributes. Many of the methods discussed here may nevertheless apply to studies of document readability more generally. By moving away from printed, and therefore fixed, text, readability is being fundamentally changed by digital text and the customization options that come with it. Readers - young and old, proficient and struggling - all stand to benefit immensely from having control over the format of the text that they read. Rather than a one-size-fits all, there is an opportunity to individuate for the reader and the reading context.

This new direction is a green field and needs new methodologies, tools, and approaches. The authors have been building these tools and methods over the last few years and will use this paper to share them with the greater interdisciplinary research community. In this pivotal moment, the future of readability is in the hands of a small group of present-day stakeholders. This paper is intended as a practical foundational resource for growing the readability stakeholder community. The subject matter requires an integrated interdisciplinary approach, and correspondingly, the authors of this paper include vision scientists, technology experts, educators, designers, typographers, and data scientists; together, we represent voices from academia, tech industry, and non-profit institutions.

This paper is written for anyone interested in pushing the state of readability research: scientists, practitioners, educators, tech companies, type designers, policymakers, and engineers. Our goal is to share our methods and tools with the greater community in hopes that together we can advance the state of the research and make readability better for all.

# 2. Reading & Readability

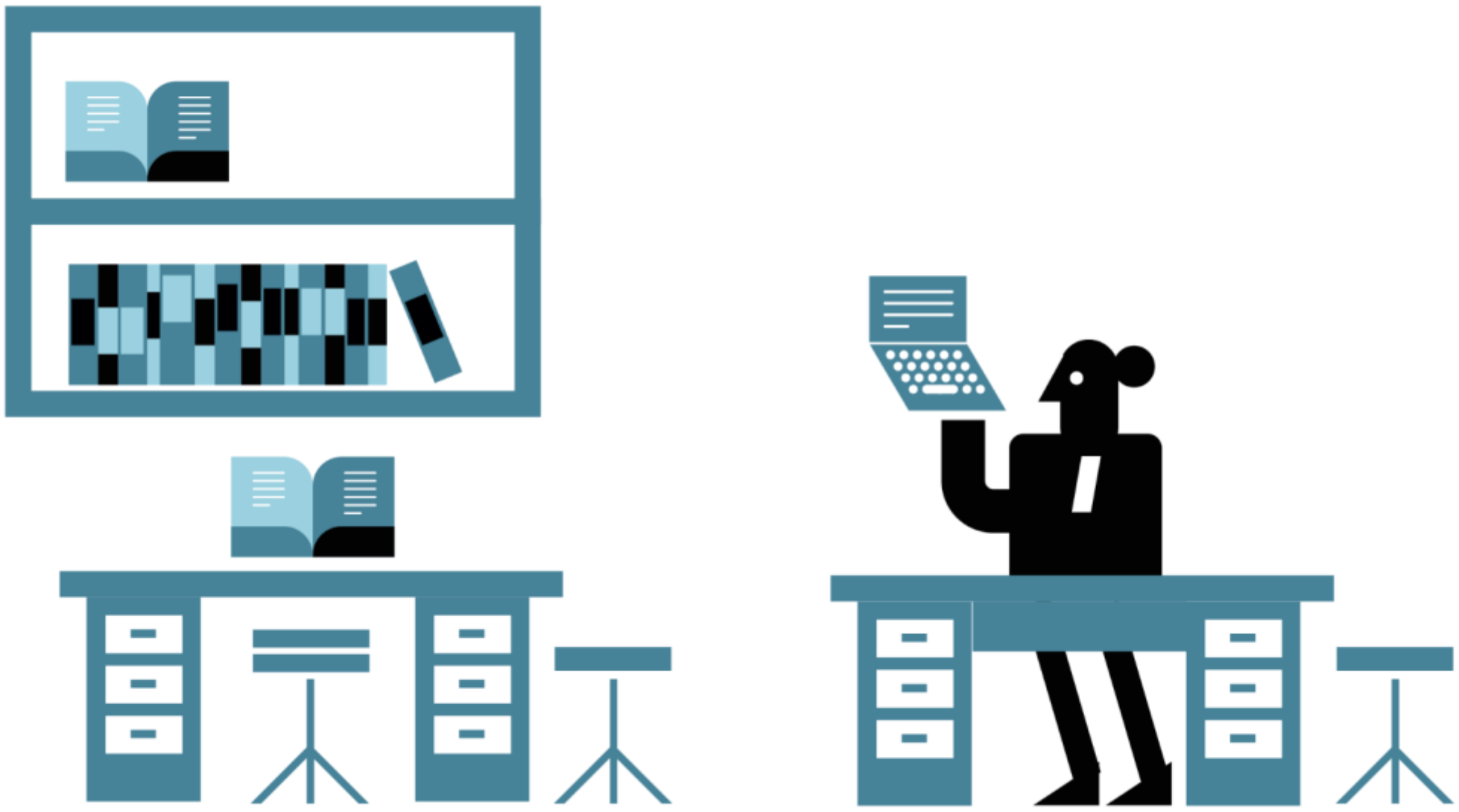

**Figure 3.** Reading is something that humans do, while readability is a property of the document we read.

The Oxford English Dictionary defines readability as “the quality of being legible or decipherable”. Readability (See Fig. 3) is informed by content, typographical features, and document-level aspects, among other unexplored properties. A change to the document’s readability affects the ease with which a reader can succeed at extracting the information they need. Optimal readability entails a fit between document, reader, and context, producing better reading outcomes.

By focusing on the presentation and legibility of text, readability research is thereby intertwined with the act of reading itself. It is essential for readability researchers to understand the core tenets of the process of reading and how these two research fields interact. As readability is a narrower and less studied topic than reading, we will occasionally borrow methodology from more general reading studies throughout this paper.

Reading is inherently a social action. We are not born knowing how to read, and we must continue to read to make sense of the world around us. Reading includes the acts of deciphering, processing, and making meaning of text and may look very different depending on when, how, and why we are reading. Considering reading, and by extension readability, means considering everything from how text is presented to the physical process of reading to the strategies that readers use while reading and how those strategies change based on readers' task and motivation.

## 2.1. Types of reading

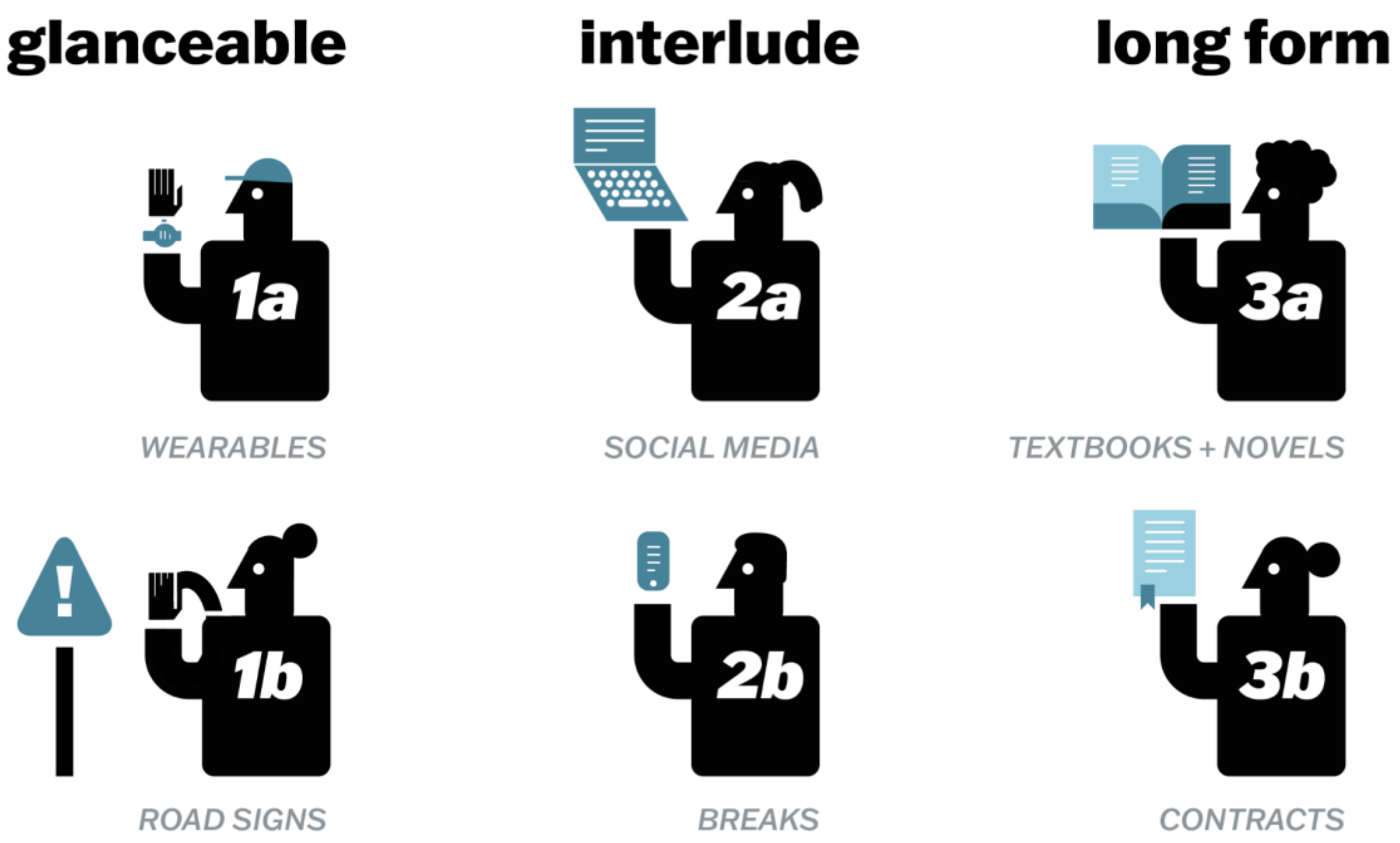

**Figure 4.** In glanceable reading, reading is not the primary task, and undue attention on reading is detrimental. In long form reading, reading is the primary task, and undue attention on other tasks is detrimental to reading. Interlude reading subtenants the complex space between, in which reading is one of many interleaved tasks.

Reading can encompass encountering a single word as we move through the world, like "stop" on a sign to engaging in more complex tasks like reading an instruction manual for assembling a new piece of furniture sentence-by-sentence, pausing as we complete each step along the way. Finally, reading can extend to ingesting large amounts of text in a single, uninterrupted pass for pleasure or knowledge. All of these reading activities, therefore, occur on two axes of intention: time commitment and purpose on the part of the reader (Fig. 4); that is, they cannot be thought of as only the simple process of deciphering text. As we consider how we design and distribute text for readability, it is valuable for us to distinguish even further these ways that people read so that we are clear about how our design goals might align readability with intentions for reading.

### 2.1.1 Reading on the Temporal Axis

We can consider reading to be a process that the reader does over time, with scanning or searching a body of text (or an interface) for a given word at one end of the reading task space. In this kind of ***glanceable* reading** (a single fixation on a single word, see Sawyer et al., 2020), a reader is reading only that one word they are looking for, rather than reading or paying attention to the interstitial text. This is the type of reading that is involved when processing the alerts on smart watches, or similar small-screen devices, or during driving, noticing the text on road signs.

Directed or guided search behavior might, then, lead to skimming the content of the document to gain context. The reader might be skimming a text to determine whether the section they are in or the document as a whole are useful to them. These skimming and searching behaviors can be considered a form of ***interlude reading*** (Wallace et al., 2020; 2020a), where readers read up to multiple sentences or a shorter part of a longer document. This is also the type of reading we engage in when browsing short snippets of news articles, headlines, or social media posts.

Moving to more in-depth reading, a reader who requires a more complete understanding will read most, if not all, words and will move progressively through a document, reading for comprehension and, potentially, for deeper meaning. According to Wolf (2016), we can define “deep reading” as reading processes, which:

> “*underlie* our abilities to find, reflect, and potentially expand upon *what matters* when we read. They represent the full sum of the cognitive, perceptual, and affective processes that prepare readers to apprehend, grasp, and assimilate the essence of what is read” (p. 112).

Deeper reading acts, such as what Wolf describes, will inherently require engaging in ***long-form reading*** (Seaboyer & Barnett, 2019), where paragraphs and pages are focused on without interruption. The reading of academic articles, textbooks, technical manuals, and legal documents all fall into this category.

### 2.1.2 Reading on the Purpose Axis

Considering reading from the point of view of what the reader *needs* brings us to reading beyond basic literacy and to the question of reading strategies and motivations. Commonly, when we think about supporting readers, we think first about basic literacy acquisition: how do we help learners sound out letters and process phonemes? This is an important first step, of course, but an ability to read goes beyond basic literacy acquisition. ***Literacy***, in other words, does not just describe an ability to recognize visual patterns in language and make meaning of them, but rather to apply that visual knowledge to social and cultural meaning-making. Such reading acts include:

1. When we read for *content uptake*, we are going beyond decoded information to learn about processes, key terms, concepts, or definitions. We might read for content uptake when we are trying to learn a new term, skill, or idea. For instance, a home cook will read a recipe for content uptake to make a meal (Wolf 2016).
2. When we read *rhetorically,* we are reading for an understanding about the context of the written work itself, its purpose and audience, and what it seeks to communicate. For example, being a critical consumer of news and social media posts requires rhetorical reading, and schools, particularly middle and high schools, have been encouraged to teach these reading skills directly so that students are prepared to be informed consumers of such media (Sweeney 2018, Brandt 1990, Haas and Flower 1988).
3. When we read for *research,* we are reading to collect ideas to create and support a larger body of knowledge, often aggregating a variety of perspectives and extrapolating themes, patterns, and ideas. A financial analyst seeking to make a prediction will engage in this type of reading (Jamieson 2013, Downs 2010, Bizup 2008).
4. When we read for *analysis,* we are reading to consider broader themes, ideas, or patterns. Often referred to as “close reading” in areas of literary study, reading for analysis involves examining readings at the word, sentence, and/or paragraph-level to make sense of how a piece of reading might fit into broader cultural or historical narratives. This type of reading is done most frequently by students of all ages and scholars in history and literary studies (Fang 2016, Fisher and Frey 2014).

These purposes for reading then depend on a variety of ***reading strategies*** (Carillo 2017, Petrosky and Bartholomae 2010). Reading for content uptake, for example, requires strategies related to information retrieval, summarization, and, depending on the goal of the reading act itself, memorization. Reading for analysis, on the other hand, may require strategies related to critical thinking and text contextualization or historicization and an ability to think and imagine meanings beyond the literal space of the text itself. Explicating the many reading strategies for approaching multiple ways of reading has been the topic of many textbooks and scholarly publications.

Ultimately, while much can be learned about readability by studying how we read as a temporal process, a vast amount is missed when the question of reading purpose and strategy is not also considered. Consider the reader: Is the length or type of material they are reading implicitly asking them to employ certain strategies? Is the timing of the study supportive of the reading purpose inherent in the task? Is the presentation of the material appropriate to employing the right strategies? Ultimately, how might these factors interact with our reader’s observable behavior, and what might be lost by not considering them? This brings us to the next section, where we focus on the reader.

# 3. The Readers

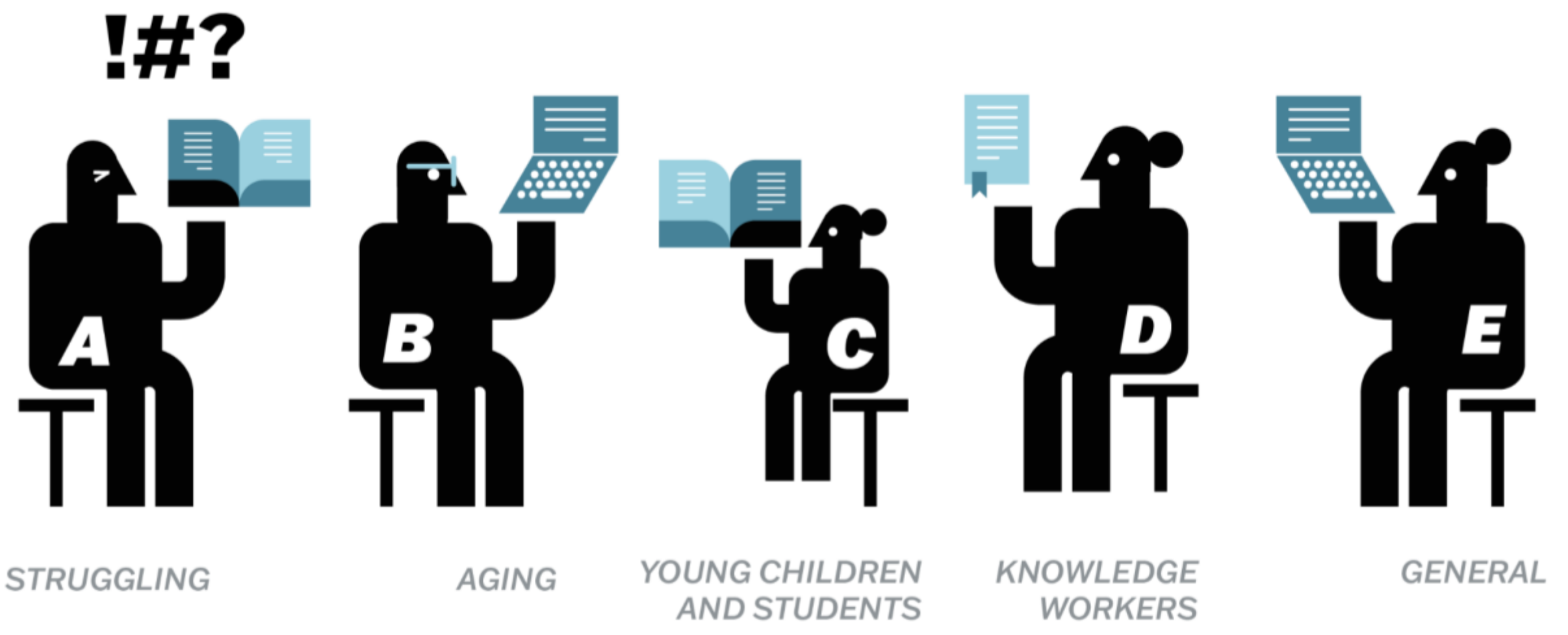

**Figure 5.** There are no outlier populations in readability; rather, this work addresses a continuum of need and a continuum of skill, in which each individual could be provided with affordances.

Readability research is first and foremost about the readers - it asks questions about the ease with which a reader can successfully decode the document (i.e., decipher, process, and make meaning of the text read). Researchers interested in particular populations (e.g., children, elderly, those with visual impairments, financial workers, cyber defenders, readers with dyslexia) should bear in mind each group's needs and abilities, whether that is in the context of compensation (and motivation), fatigue, or their ability to do the task of interest. In this section, we discuss how to design a readability study with the attributes and constraints of the target reader population in mind, as well as considerations that apply in different settings, for example, running the study in the lab or in the wild.  We conclude this section with a discussion of the foundational ethical research principles that apply to working with human participants in any study. These include the need for participants to be part of the study willingly, for their privacy to be respected, and for the benefits of the research to be commensurate with the risks involved.

## 3.1. "Who" are the readers

Readability affects any literate or nearly literate child or adult and spans multiple expertise domains. This means that nearly any reader can help us understand how they read and how they might read more easily and more completely by acting as a participant in reading studies.

**Struggling readers.** In our view, there are no outlier populations; rather, participants should be seen as lying on a continuum of need and on a continuum of skill (Shaywitz & Shaywitz, 2020).

Nonetheless, psychology and many other disciplines focus upon deficits, and indeed investigating reading in people who struggle with reading can help us understand the cognitive, linguistic, and environmental factors that influence how we read, which could in turn affect readability. One population of particular interest to many reading researchers is individuals with developmental ***dyslexia***. Given the complex nature of reading that involves the visual, phonological, and semantic processing of the reading materials, it is increasingly clear that dyslexia is a multifactorial condition (Pennington, 2006; Ramus et al. 2003; Vellutino et al. 2004). One of the most popular theories is the phonological impairment theory of dyslexia. However, the extent to which phonological awareness specifically contributes to reading deficits is still debated (Catts et al., 2017; Fostick & Revah, 2018). Similarly, whether dyslexic readers have elevated visual crowding or poor motion sensitivity compared to non-dyslexic readers is also not fully understood (e.g. Joo et al., 2018; Demb et al., 1995; Olulade et al., 2013; Martelli et al. 2009). Regardless, studying this population has helped us understand how visual, phonological, and semantic processing contributes to reading, which will help us understand how to improve readability for all readers. Another population that can help us understand readability more deeply are non-native speakers who may have deficits in oral language and reading comprehension skills, despite adequate decoding capabilities (Spencer M. & Wagner W. 2017).

**Aging readers.** People with declining reading skills may similarly appreciate improvements to their reading experience to compensate for the loss of visual acuity (e.g., age-related presbyopia), declining cognitive ability (Bokulich et al., 2016; Owsley, 2011), and changes to other sensory capabilities such as crowding and visual span (Rayner et al. 2010; Legge et al., 2007; Levi, 2008). Reading speed is known to slow down with age (after age 40), while reading acuity decreases and, as a result, critical print size increases (Calabrèse et al., 2016). Interestingly, beyond population-level opportunities, individuated font and other typographical characteristics of a text can likewise drive increases in readability (Wallace et al., 2020; 2020a).

**Readers learning to read.** Populations who are in the thick of learning to read can also benefit from reading interventions (Powell & Trice, 2020; Reid et al., 2004; Hughes & Wilkins, 2002) including older readers not fully proficient in early literacy skills such as letter knowledge and print awareness (e.g., adult literacy learners) (Graesser et al., 2019; Sabatini et al, 2011). In particular, the learning-to-read population of young children roughly between the ages of 3-7 has not been very well studied in the context of readability and its effects on the formation of sound-letter understanding. An increased understanding of how readability impacts this critical learning process has the potential for great impact, as research has shown that when students are not proficient readers by fourth grade, they are far less likely to complete high school with serious consequences for economic and civic prospects for the remainder of their lives (Cramer et al., 2014; Hernandez & Napierala, 2013).

**Knowledge workers.** Other points on the continuums of need and skill are the knowledge workers—including medical experts leafing through patient records (Nygren et al., 1992; van Engen-Verheul et al., 2016; Bouaud & Séroussi et al., 1996; Elson & Connelly, 1995; Henriksen et al., 2020), cybersecurity experts scrutinizing immense amounts of text content for potential threats (Lotem  et al., 2012; Macy et al., 2014; Lee et al., 2019; Ehrlich et al., 2017; Jang-Jaccard & Nepal, 2014), financial analysts integrating information from various sources for making predictions (Lehavy et al., 2011; Li, 2010; Loughran & McDonald, 2010; Ravula, 2021; Hoitash et al., 2021; Bradshaw, 2011), and scientists immersing themselves in academic reading to stay up to date (Eung et al., 2018). These populations of heavy readers have the potential of experiencing immense career benefits from tools that improve the throughput or quality of their reading.

**The "general" reader.** Readers in a population span a myriad continua of need and skill, and perform the task of reading in a multitude of contexts. The 'general' reader can therefore be seen as the reader that reads regularly for pleasure - e-mails, social media feeds, news, interest articles, blogs, etc. - and is frequently engaged in interlude reading of a few sentences or paragraphs at a time. Among the readers without any specifically diagnosed reading or learning difficulties, we still encounter a wide range of reading speeds and abilities, from less-versed readers reading below 180 WPM to speed readers capable of speeds up to 750 WPM (Rayner, 2016). It is often assumed that average adult readers read between 200 and 250 words per minute. A study by Brysbaert 2019) found that the average silent reading rate for adults in English is 238 words per minute for non-fiction and 260 for fiction. Moreover, however, it is advantageous to focus studies of readability to a particular population, and perhaps refine further by a particular context.

Researchers, engineers, designers, and others interested in developing tools for improving readability need to study reading across many populations, by considering diversity in age, reading skill level, pre-existing deficits, and other participant characteristics. This will help develop individualized recommendations as well as a better understanding of the wide variety of readability factors that affect some individuals but not others. From a study design perspective, different populations require different considerations. For instance, children, elderly adults, individuals with reading, cognitive, language, or neurological disorders may fare better under shorter study sessions. In such cases, multiple short data collection sessions rather than a single long session can often yield better quality data (see Section 6).

Once a population of interest has been identified, attention must be paid to population features that may influence experimental results if not properly controlled. Aside from the commonplace factors of age, education level, and occupation, additional sources of variability can result from whether vision correction is used (and whether it is used during the study), the participant's reading proficiency (e.g., reading level), prior diagnoses of reading or learning disabilities, any eye conditions (e.g., cataracts, glaucoma, retinal degradation), possible influences of stimulants or depressants (including common ones like nicotine and caffeine), other languages spoken/read by the participant, lighting conditions, and reading environment at time of study, etc. These features can be tracked with study surveys (either as a pre-screen or post-completion questionnaire) and then used either for participant filtering or incorporated as factors in the analysis. Since there are many potential factors to take into account here, and only some will apply to any given population or study, we provide a sample demographic survey in the appendix to this article as a starting point.

## 3.2. “How much” population data is enough

With the particular population of readers pinned down, the researcher will next need to decide how many readers to recruit for a given readability study. Of course, the degree to which researchers can generalize from their data is a function of both the population *and* the question; no one size fits all situations.

**Size matters.** Size depends not just on the raw number of participants, but also on the number of trials per participant, the total amount of data gathered, and the diversity of the participant pool - in its ability to act as a representative sample of the target population. A good way to think about this is not “how many participants” but how many “experimental units” are available for downstream analysis. For instance, it is not uncommon for behavioral/neuroscience studies to derive conclusions from studies of one or two dozen participants (Sihoe, 2015). Data from such “low-N” studies needs to be very high quality, with possible confounding factors identified and controlled for, and many trials run per participant to ensure the trends captured are representative of an underlying truth, and not simply of spurious factors; confounds in design, outliers in the population measured, unforeseen external factors, and so on... On the contrary, when running large-scale studies of hundreds or thousands of participants, on crowdsourcing platforms for instance, a large N can help wash out individual participant noise (Bolthausen & Wüthrich, 2013). The right N depends in part on the key study questions - i.e. the metrics used, practical significance level desired, statistical power (Lan & de Mets, 1994; Per Broberg, 2013), number of variables in the experiment, and the nature of variables and experimental design. Prior readability studies (Banerjee et al. 2011; Bernard et al., 2001; Boyarski 1998) have made general recommendations on the basis of dozens of participants’ worth of data, but these recommendations can change dramatically as the number of participants is increased by an order of magnitude (Wallace et al., 2020b).

**Individuals matter too.** For painting a picture of the variability inherent to a population, or with a focus on individuation, fewer participants is an acceptable starting point. In this case, the benefit of large numbers is to contextualize the behavior of the few - are they representative of participant clusters, or outliers in the data? So while size matters, zooming in on individual participants can help paint the stories of real impact and change (Crowley & Jordan, 2019). Indeed, recent work on individuation in readability suggests that significant gains are available by looking beyond measure of central tendency, and to clustering and individual differences approaches (see Wallace et al, 2020, 2020a).

## 3.3. “How” to recruit

**First and foremost, treat participants ethically.** Recruiting participants is a big challenge for capturing an accurate sample of high-quality, real-world data, and it is essential that researchers treat participants respectfully and ethically. As a result, participants should be given multiple opportunities to consent, and the researchers must clearly explain any risks so the potential participant can make an informed decision. If research is done unethically, it can put participants at risk and damage the trust between researchers and participants more generally (see also Section 3.5).

**Partner with a domain expert.** Working with a domain expert can simultaneously simplify the recruiting process and provide important insights into the attributes and limitations of the target population. For example, partnering with a school district, individual educator, or third-party reading organization can provide (i) access to student populations, (ii) requirements of working with minors (e.g., participant assent, parent permission slips; more details in Section 3.5), and (iii) guidance for tailoring the directions, tasks, and content of a study to the appropriate comprehension level (e.g., depending on the presence or absence of an adult supervisor or teacher). One can approach organizations that directly work with adults developing literacy skills or second language learners (e.g., libraries, community-based organizations, and federally-funded adult learning programs). Additionally, networks of these adult learning organizations might be found through larger nonprofit organizations like World Education, Inc., ProLiteracy, and COABE. One can also approach organizations that support K-12 students and educators (e.g., curriculum providers and after-school programs).

**Consider the advantages and downsides of crowdsourcing platforms.** If it makes sense for the study question to consider an easy-to-recruit sample of the general population, crowdsourcing platforms like Amazon's Mechanical Turk, UserTesting.com, Crowdflower, and Prolific are available (Buhrmester, et al. 2016; Paolacci & Chandler, 2014; Peer et al. 2017). These platforms provide a wide range of readily available users motivated extrinsically through monetary compensation. Other models such as friend-sourcing are where someone relies on recruiting their friends to participate voluntarily, which could significantly bias results but provide an easy platform for pilot testing experiments. A cheaper alternative to paid crowdsourcing that still provides some of its benefits is relying on volunteers. LabInTheWild, for example, is a citizen science platform that has been able to tap into users' intrinsic and altruistic motivations to recruit thousands of users to perform various studies (Reinecke et al. 2015), including a reading study that showed that a special "Reader View" in web browsers could increase reading speeds and improve the perceived readability and visual appeal of the text (Li et al, 2019). By providing users insights on how they compare with others in terms of reading speed, preference, and other reading tasks, participants are motivated to share the studies with others and to return for additional studies.

Special populations can also be recruited via targeted messages in relevant forums such as Reddit (Shatz, 2017) or by advertising in social media or markets like Craigslist (Alto et al. 2018). The relevant considerations in all these cases are the potential for self-selection (e.g., participants who are more likely to volunteer for a reading experiment may not be representative of all readers) and the impacts of incentives on performance (Mason et al. 2010; Morris et al. 2010; Ho et al. 2015; Morris et al. 2017). Our own experience is that having a diverse research team representing different spheres (e.g., scientists, educators, technologists, designers) can help with participant recruiting and designing a study with greater awareness of the attributes and limitations of the target population (e.g., their interests, prior knowledge, attention span, available resources, as well as physiological, psychological, and social constraints).

## 3.4. “Where” to conduct studies

Tightly intertwined with the choice of target population are considerations about where the readability study may take place. While the features of the population may provide constraints for the study location, often the experimenter is still faced with a number of choices (e.g., whether to conduct a study on adult readers in a controlled in-lab environment or online; whether to conduct a study on children in the classroom or the home, etc.). Each choice of location correspondingly involves trade-offs between ease of recruiting, data quality, and ecological validity of the reading environment.

### 3.4.1. Laboratory-based in-person studies

An important consideration for choosing where to conduct a study is the obtainability, accessibility, representativeness, and reliability of the participant data. From this perspective, in-person, lab studies offer the highest degree of control and access to participant reading data. Moreover, the experimenter has the option of taking physiological recordings of the participants while they read - via eye trackers, brain imaging technology, or other sensors that can be set up and carefully calibrated for each participant (see Section 5).  For fundamental questions in readability, including many questions about the visual mechanics of reading (how readers move their eyes and why, depending on task, goal, experience, and strategy), laboratory studies are extremely revealing. However, there is always the tension between results from laboratory studies and ecological validity - i.e., the behavior that readers engage in outside the laboratory. After all, research labs involve performing unfamiliar tasks in unfamiliar settings for most participants. While some of these gaps can be overcome by replicating study designs outside the laboratory, or by developing and using more naturalistic tasks in the laboratory, this gap will always exist, and minimizing it is a function of developing fundamentally generalizable tasks that reveal the underlying mechanics and processes of reading, where those findings can be applied to behavior outside the lab.

### 3.4.2. Remote studies online

The main advantages of conducting remote studies - e.g., studies hosted online on crowdsourcing platforms - are the ease, speed, and available quantity of participants. In particular, the numbers of participants available online can be substantially higher than those that are able to come in for in-person studies. As discussed in Section 3.1, a larger sample size can produce a more robust statistical distribution and easier-to-spot outliers. While remote studies can increase the diversity of the participant pool, by not being geographically limited, it is important to realize that these platforms self-select for participants who own a computer or phone and volunteer to participate in such studies. With society's migration to working remotely instead of in-person due to an ever-changing world, conducting remote studies is also growing in prevalence and acceptability.

**Beware of unobserved reading behavior.** When using measurements captured from the reading behavior of online participants, or in situations where the participant is otherwise not directly observed, there may be numerous variables that could impact the readings, including distraction or multitasking (Ophir et al., 2009; Reeves et al., 2020), task switching, or "short-cutting" activities like taking screenshots to facilitate comprehension or memorization tasks. While crowdsourced data from voluntary participants in their homes heavily increases the obtainability of the data, such data must be handled with care and analyzed for anomalies that should be reported and justified as exclusions during analysis (Section 7.2). This is because the data comes from experiments done at home with more uncontrollable variables, and participants may lack the necessary training to adjust for this accordingly.

**Capturing reading in-the-wild.** While remote studies might lose some internal validity by giving up control of the reading environment, they gain ecological validity by studying participant reading habits in their natural environments, usually from the comfort of their homes. Recent work has introduced novel methods to control for participants' environments in remote studies. One example is the "Virtual Chinrest", a method that measures a participant's viewing distance from the screen in the web browser (Li et al. 2020). Using the Virtual Chinrest, the researchers were reliably able to measure ***visual crowding*** in an otherwise uncontrolled online environment. Visual crowding depends on a precise calculation of **visual angle** and has been shown to affect people's reading performance (Joo et al. 2018). This method provides a promising pathway to web-based reading studies that require precisely controlled stimulus. There has also been work on capturing eye movements remotely using web or cell phone cameras (Papoutsaki et al. 2017), allowing both for an additional measurement of observer behavior, but also providing validation of whether study participants complete the task honestly - i.e., actually moving their eyes to read the content on the screen instead of simply clicking through the task. To increase internal motivation for completing the studies, many researchers in other fields gamify their online studies or provide personalized feedback, leveraging the general growing interest in personal informatics and self-experimentation. Readability studies can similarly offer personalized font or reading format recommendations, motivating readers with possible improvements to the user experience or reading effectiveness (Wallace et al. 2020).

### 3.4.3. Remote studies in-context

Another option for running studies outside of the laboratory is by partnering with organizations or professionals with access to specific reader populations, including school-age children or knowledge workers. Such studies can stand to benefit both from (1) external validity, by studying participants in their usual reading contexts and (2) control over, and reliability of, the data capture, since the collaborating organization or individuals may help administer the study and directly observe study participants.

As a case in point, partnering with educators to conduct both small- and large-scale studies can be an effective method to evaluate the impact of readability on reading outcomes for learners of all ages. Teachers, and the learners themselves, can provide additional insights, quantitative and qualitative, which can assist in assessing impacts, and, like with remote studies, classroom studies gain external validity by studying participants' reading habits in their natural environments. There are important ethical considerations with classroom-based research, which we discuss in Section 3.5.

In professional contexts, one may choose to study populations from the military, healthcare, and financial institutions that engage in reading as part of their job (see also Section 3.1). In these cases, the corresponding readability study may leverage the reading materials that would be familiar to the study participants, provided there are no privacy or security risks. It is also important to consider potential risks related to either a readability study, or resulting recommendations, inadvertently interfering with the performance or accuracy of the professionals in their jobs.

### 3.4.4. Remote studies on-the-go

With the screens of mobile phones increasing in both size and resolution, reading has become much more attractive on these devices. Their pervasiveness allows users to engage in reading sessions while on-the-go, e.g., during the daily commute, in waiting situations, at home as well as at work. Unlike when reading on desktop computers, the context of mobile phones can widely vary in terms of environmental conditions (e.g., lighting or noise levels) and users' primary or secondary activities (e.g., walking). Research by Liu (2005) has shown that reading on mobile devices has led to users engaging in rather brief reading sessions characterized by skimming behaviour as opposed to in-depth reading sessions. Oftentimes, the reading activity is not the primary user task as people interact with their phones while walking or during conversations. Such additional tasks introduce higher workload and, therefore, influence the reading performance (Schildbach, 2010). Designers can rarely rely on users' undivided attention and also need to expect sudden interruptions to the reading task.

Studies on-the-go strive to investigate realistic situations, in which users turn to their devices to read, as studying reading in naturalistic environments, as noted in Section 3.4.3, can greatly increase ecological validity. Contextual conditions are generally not controlled however, which makes interpretation of reading behaviour challenging. Insights into the context of particular reading sessions can be gained by collecting phone sensor data or using experience sampling (Van Berkel, 2017). Data collection frameworks, such as the AWARE framework (Ferreira, 2015), combined with activity recognition algorithms (Krumm, 2010) allow investigators to make sense of the context of reading sessions on-the-go.

Reading on-the-go can also be studied under controlled conditions by introducing secondary user tasks, such as walking. So-called in-motion scenarios are often more representative of mobile phone usage but require the conscious selection of specific evaluation methods. Consider the use of a controlled walking scenario, such as a pre-configured walking trail (for example, a figure-of-eight in confined spaces), or the use of a treadmill can recreate a realistic on-the-go scenario in the lab. Of course, a brief review of the literature can help the cautious researcher to avoid methods with undesired limitations: Barnard et al. (Barnard, 2005) found that treadmill walking allows experimenters to analyze reading performance aspects, but it is insufficient to identify contextual factors. For more realistic measures of performance, accuracy and workload, a walking scenario is preferable. Performing controlled studies in real-world scenarios can be challenging as studies by Kjeldskov and Stage (Kjeldskov, 2004) show: by comparing in-lab evaluations of mobile devices in comparison to real-world settings, they show that laboratory studies provide a good approximation for the user performance but often fail to capture the user's additional workload realistically. Complex contexts are seldom well-represented in simulations, and it is up to the researcher to determine how reading-on-the-go experimentation balances between practicality and safety on one side, and generalizability on the other.

### 3.4.5. Replicating studies across environments

Important questions for readability research include (1) to which extent reading behaviors replicate across different environments and (2) for the effects that don't replicate, which factors drove those effects in the first place? With remote studies, we can reach larger numbers of people to understand individual differences. However, these results might compromise some internal validity in exchange for external validity. Rerunning the studies in controlled environments in a lab might tease apart which observations are reproducible across different environments. Replicating results from a remote environment in an in-lab environment will add the missing internal validity to remote studies. On the contrary, re-running laboratory studies with remote volunteers or paid crowdworkers can help researchers understand if the results they observe exist in more natural behavior or only in controlled laboratory settings. There is also utility in comparing results in and outside of the classroom. While teachers can provide qualitative observations of students' reading behaviors in class, how do the same students perform at home, with less teacher oversight and possibly increased distractions? Do the readability formats that work best in carefully controlled laboratory environments generalize to natural classroom environments, filled at times with distractions and interruptions?

## 3.5. Oversight

The question of ***research oversight*** is how we balance our desires as readability researchers, who want to collect data to better understand readability, with the rights of our participants. This oversight can take many forms and varies considerably (by setting and country), so we will discuss universal fundamentals of research oversight; that is, what any readability researcher must consider. While we might want to focus on the data that we need for a given study, it is also imperative that we consider our participants' rights in the study and their rights regarding their data. We will use the following universal principles to frame this discussion: *respect for persons*, *beneficence,* and *justice*.

Research with human participants must respect participants' choices with regard to participation, data inclusion, data retention, and protect them from harm that might occur as part of the study or thereafter. This includes the question of anonymity and potential harms that might ensue if a participants' data can be linked back to their identity. If, for example, an experiment involves a screening procedure for a disorder, and that data were made publicly accessible, it might cause harm to the participant. This is a particular concern with data that are difficult to truly anonymize, such as video or audio of participants. As researchers, it is our responsibility to ask our participants to provide informed consent before they participate and that this consent is based on their understanding of what they will be doing, the data they will generate, and what we will do with that data long-term.

The principles of beneficence and justice in research make it imperative that we do our best to minimize potential harms and maximize the benefit of our research. That does not mean that a given participant must derive personal, immediate benefit from participating, but rather, by participating, they will help us increase our knowledge. This includes fundamental research - increasing our understanding of the basic mechanisms of reading will benefit society because it will help us improve the reading experience for all readers. Justice in a research context is the question of who bears the burdens and reaps the benefits; ethical research cannot only study one population and never return the fruits of that research to them. In readability research, it means that we must include a focus on disseminating our results and their implications. If our goal is to understand reading more completely to help everyone, and we use the products of our research to move towards that goal, we integrate the principle of justice.

These are broad, bedrock principles for ethical research - each research environment, whether that is research in university settings with undergraduate students, in-classroom work with school-age students, *in situ* research with professionals, or online research has its own constraints and applicable regulation, and complying with them is the researcher's duty and responsibility.

Research in school-age settings where the participants are minors brings its own issues, including the need for parents to provide informed consent (while the minor participant provides assent) and the consent of the administration and other students - centering the principle we have discussed of respect for subjects and the idea of minimizing harm. It is also imperative in these settings to hold data confidential and anonymous because the risks are magnified with some populations and some data (particularly video or audio of children). In addition, the principles of beneficence and justice come into play here because researchers must balance the potential benefit of the data they collect with the potential harm that their disruption of the classroom may cause for other students. Critically, withholding an intervention in order to have a control group is not ethical if it adversely affects the control group's expected educational experience.

Considering a different example, online research, whether it is explicitly experimental (e.g., with crowdworkers) or manipulations integrated into a platform, service, or application, must reflect these same principles for ethical research. It is not enough to merely ask what might happen, manipulate a user's experience of service and collect data on their behavior to increase profits. Research with participants or users needs to center the participants (or users) to reflect the need to respect them as people and as participants.

On the whole, questions of oversight or research ethics should not be thought of as constraints to researchers in academic or private sector settings, but rather as a framework meant to keep everyone safe. Studying readability and focusing on improving reading is best done, we argue, by doing it ethically and carefully and realizing the greatest benefit for all readers, building trust and knowledge together.

# 4. Reading Materials

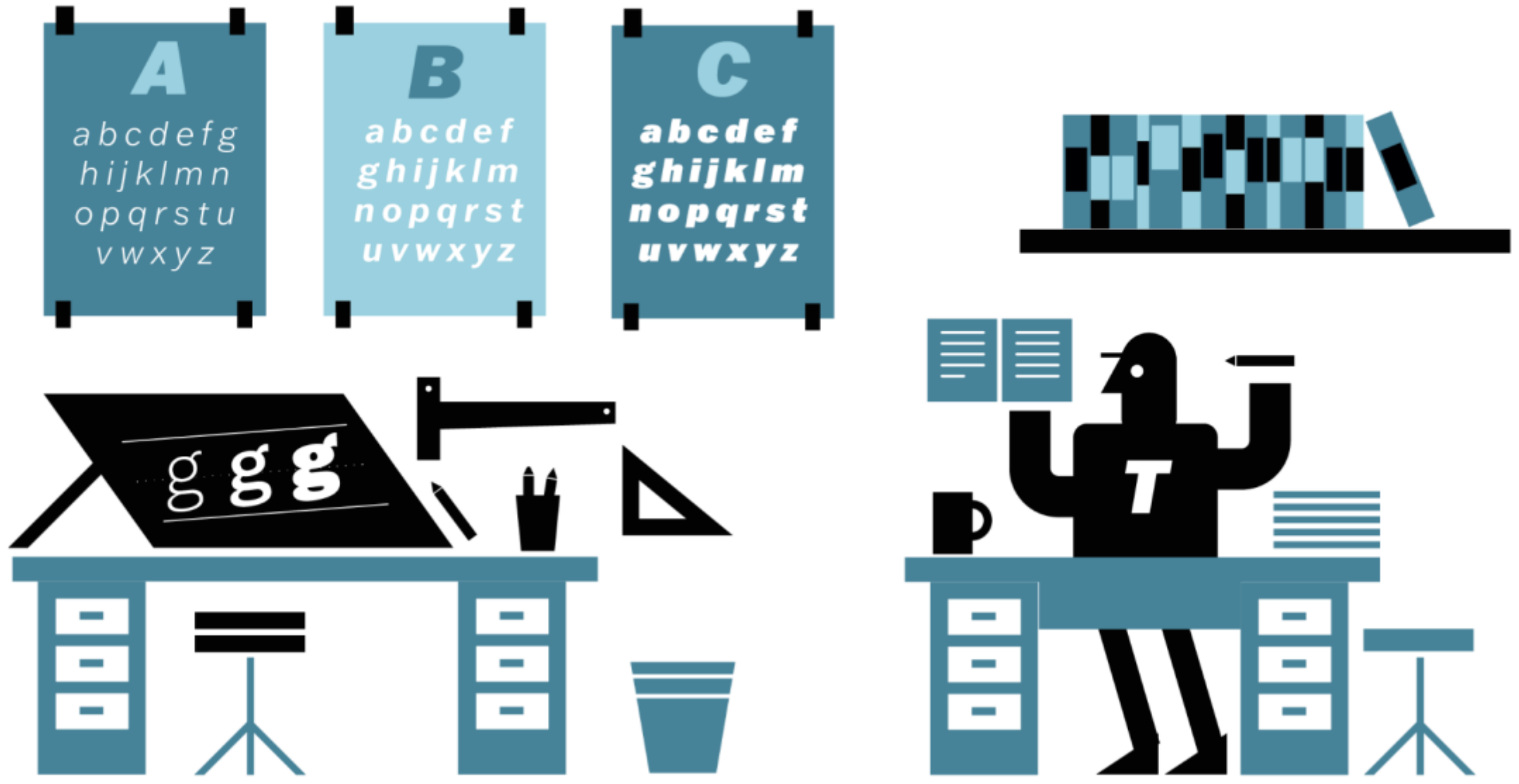

**Figure 5.** Typographic properties include the content, how it is presented, and how it is laid out.

Inextricable from when, how, and why readers read is the question of what a reader is reading: the content, how it is presented, how it is laid out (i.e., typographic properties, Fig. 5) are key to whether the content or its presentation pose difficulties to the reader as they read.

## 4.1. Content Curation and Leveling

Reading material must be curated to be appropriate in topic, length, and level to the target study population. Familiarity with the topic and interest in it are possible confounding factors (Wallace, 2020a), as both can affect the ease and speed with which the readers consume the content and whether they switch to a skimming reading mode. Length must also match the skills and abilities of the target population. Study fatigue or, worse, the inability to complete the study task can significantly affect a study's conclusions. Genre is relevant for speed and comprehension, in that, for example, narrative texts tend to be easier for readers as they have a more familiar chronological organization (Hiebert et al., 2010). Texts must also meet target populations' expectations for reading inoffensive topics so as not to distract and should not reflect bias or stereotypes about any group. For example, contentious topics may elicit reactance (Brehm, 1966), which may prompt significant cognitive engagement with the material, affecting speed, comprehension, and emotional affect. Finally, the topic of the reading material also directly affects the level of the material, particularly in the vocabulary employed, and must be appropriately tailored to the target population.

For assessing the level of a reading passage, the standard today is computer-based readability indexes. A ***readability index*** is a way of measuring the ease of comprehension of a piece of text (McCallum & Peterson, 1982). Educators traditionally designed formulas to calculate readability manually (McCallum & Peterson, 1982; G. Klare, 1976), but researchers have since introduced several methods of analyzing text with automatically-computed indexes. Some of the most common methods used today include: Flesch Kincaid Grade Level, Flesch Kincaid Reading Ease, Gunning Fog Index, SMOG Index, Coleman Liau Index, and the Automated Reading Index (Brigo et al., 2015; Mc Laughlin, 1969; Zhou et al., 2017). These invoke mathematical equations that use measures of word difficulty such as average word length and word frequency (e.g., Mean Log Word Frequency or MLWF), sentence length and syntactic consistency (e.g., Mean Sentence Length or MSL), and passage length (calculated using word counts, syllable counts, or character counts) to make predictions about the reading level of the passage. All these features of the text interact and cannot be viewed in isolation. For instance, different calculators vary in response to multisyllabic words, possessive nouns, symbol choice, and symbol pronunciation and syllable count. (Zhou et al., 2017). Readability calculators are generally reliable for ordering text into levels and predicting the rough difficulty of passages. Nevertheless, because they are not always consistent, leveling is usually carried out by considering multiple indexes simultaneously. Furthermore, some researchers advise that automatically-computed indices stop providing meaningful results at tenths of grade estimates (Zhou et al., 2017). Which indices should be used together is poorly understood and should be the focus of additional research.

Despite the frequent use of readability indices for reporting text difficulty levels in studies across a broad range of domains (Agarwal et al., 2013; Loughran & McDonald, 2014), guidance on how to precisely match readers to different reading index levels is not generally available ( K. Olson, 2010), so we recommend consulting with an education expert on questions of appropriate content levels for a given target population. For example, for both K-12 students and adult literacy learners, the text they are reading should feel relevant to them, hold their interest, and be a good fit for their literacy proficiency. The instructors you partner with can help you determine appropriate text difficulty. Additionally, adult learners need to know why they are engaged in the activity - especially if they are busy adults with families and jobs who struggle to make time for learning. They will want to know why they are taking valuable class time for the activities and will be more engaged if they do (Knowles, 1970).

**Towards open-sourced reading materials.** The presence of standardized reading corpora would both lower the barrier to running new readability studies and facilitate cross-study comparative analysis to move the field forward. Unfortunately, while some resources are available, they are scattered across communities and hard to come by. In the Appendix, we provide a list of current available reading corpora.

## 4.2. Typographic Considerations

In recent years, a growing number of interdisciplinary research groups involving both type designers and scientists have emerged. This, in combination with the possibilities for developing and controlling font stimuli with the variable font format, will open up new experimental settings where the font stimuli are much more controlled.

When presenting content for reading, one must make decisions about the fonts, sizes, weights, colors, spacings, and other visual features of the presented text. When studying readability specifically, one may be comparing different text formats to each other. In these cases, methods may need to be employed to normalize the fonts and type settings.

The following mantra has been used for font selection since the beginning of personal computing: script, language, category, (classification) typeface, font, glyph, size, color, column width, line spacing, and letter spacing. Each of these choices has an effect on reading. In addition, the presentation of these fonts will be influenced significantly by hardware and software used to present text for reading.

### 4.2.1. Understanding Font Classifications

There are several thousand written languages represented by close to a 100 modern scripts. Each of the scripts have implications for reading (Kessler and Treiman, 2015). In the Latin script taxonomy, categories like serif, sans-serif, handwriting, blackletter, etc. represent fonts based on their anatomical characteristics. Classifications are used to identify more specific anatomical features like the shape of the serif, or the angle of stress, or angle of terminal. Fonts also vary on many parameters, such as stroke modulation, letter skeleton, and letter proportions (for a deeper analysis of typeface classification see Bringhurst 2004; Tracy 1980).

**Uniquely identifying fonts.** Referring to fonts solely as “Garamond”, “Caslon” or “Bodoni” is not enough information to be able to identify the font. Digital fonts that are based on historical sources exist in multiple versions. As an example (Fig. 6), the group of Garamond typefaces is a revival of an Old Style serif design by a 16th-century French engraver, Claude Garamond.

Typefaces of this group include versions from Adobe, ITC, Ludlow, Peignot, Stempel, URW, Berthold, Monotype, etc. For the benefit of study analysis and future researchers, we recommend accurately noting the specific fonts used in a study, referencing the Family - Style and any attributes applied (e.g., CSS property setting, like “ITC Garamond - Regular”).

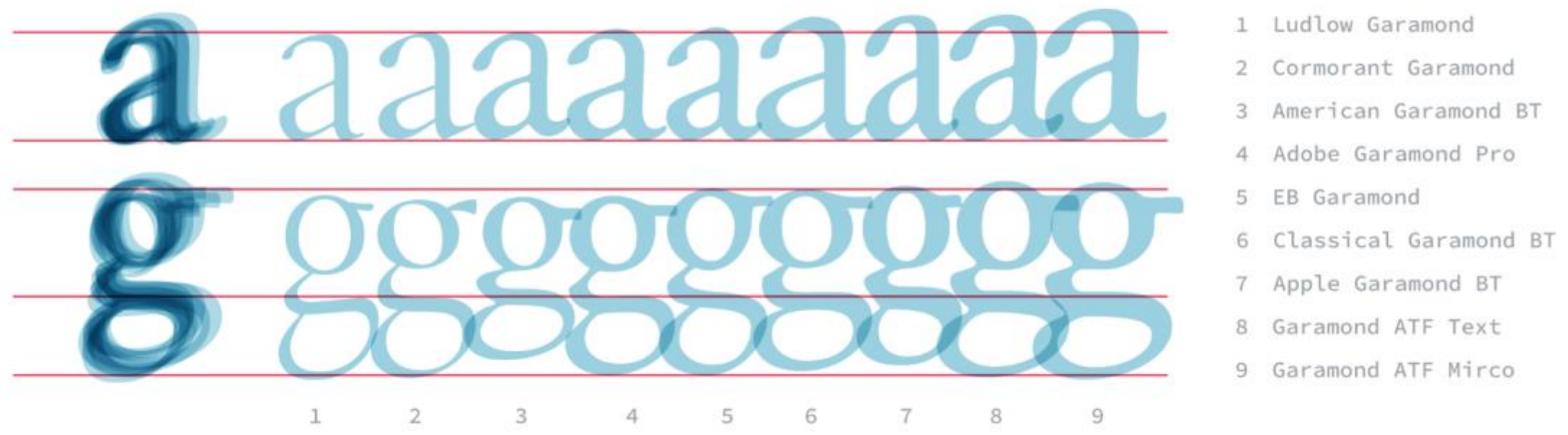

**Figure 6.** Examples of Garamond fonts from nine different foundries. Note how much they vary in the design of the letters. The letters are superimposed at the left to make the amount of variation more visible.

## 4.2.2. Selecting Fonts based on Availability

There are over 600,000+ publicly available fonts. At most, a few thousand typefaces have been designed to improve ***readability***. Only a few hundred of these typefaces have been optimized for screen, and only a few dozen are ubiquitous. This section will elaborate on various matters to consider in the visual presentation of text content for reading experiments.

For many readability researchers, the availability of test fonts is a practical consideration. Times, Arial, Georgia and Verdana are the most common typefaces, and are often used in studies (Wallace et al. 2020; Pušnik 2016; Bernard 2001). Sometimes called the ***web safe fonts***, they appear on all Apple and Microsoft products and therefore are available to all web browsers. In addition, Google, Adobe, and IBM have also made high-quality typefaces available for free distribution. At the time of writing, some of these fonts can be found on the websites of Google Fonts (https://fonts.google.com) and GitHub (https://github.com). Such fonts can be used free of charge, hosted or self-hosted. In general, fonts should be legally obtained and used with consideration of the licensing terms.

**Intentional and unintentional font replacement.** All operating systems have lists of font aliases that are used when a popular typeface is not available. For example, while Helvetica comes preinstalled on Apple’s macOS, it is not available on standard Microsoft Windows installations. Instead of Helvetica, Arial will be used. This also applies to the web safe fonts Arial, Times, Georgia, and Verdana, which are not available on Android devices and are all aliased to various fonts in the Roboto family. Users and IT administrators can also install fonts with the same name as a popular (or web safe) typeface that are completely different. Even more insidious, they could install a different version of the same font with small, but significant, design changes. Therefore, we recommend always bundling the typefaces with your test cases, to prevent unwelcome surprises while running experiments. If you can’t bundle the typefaces, test if the expected fonts are used on all devices and browsers you intend to use.

Essential decisions of how to visualise test stimuli should ideally be made by professionals. A skilled typographer not only understands the theoretical foundations of what makes a typeface function in a given reading situation, but also draws on a significant amount of tacit knowledge, acquired through years of schooling and training.

### 4.2.3. Controlling Font Properties

No matter the experimental paradigm, the visual appearance of stimuli will always affect the final results. A common approach in readability research is to compare reading performances using fonts of different typeface families. Different font categories (Figure 7) vary on multiple properties (Figure 8), anatomy, and attributes, each of which can affect reading.

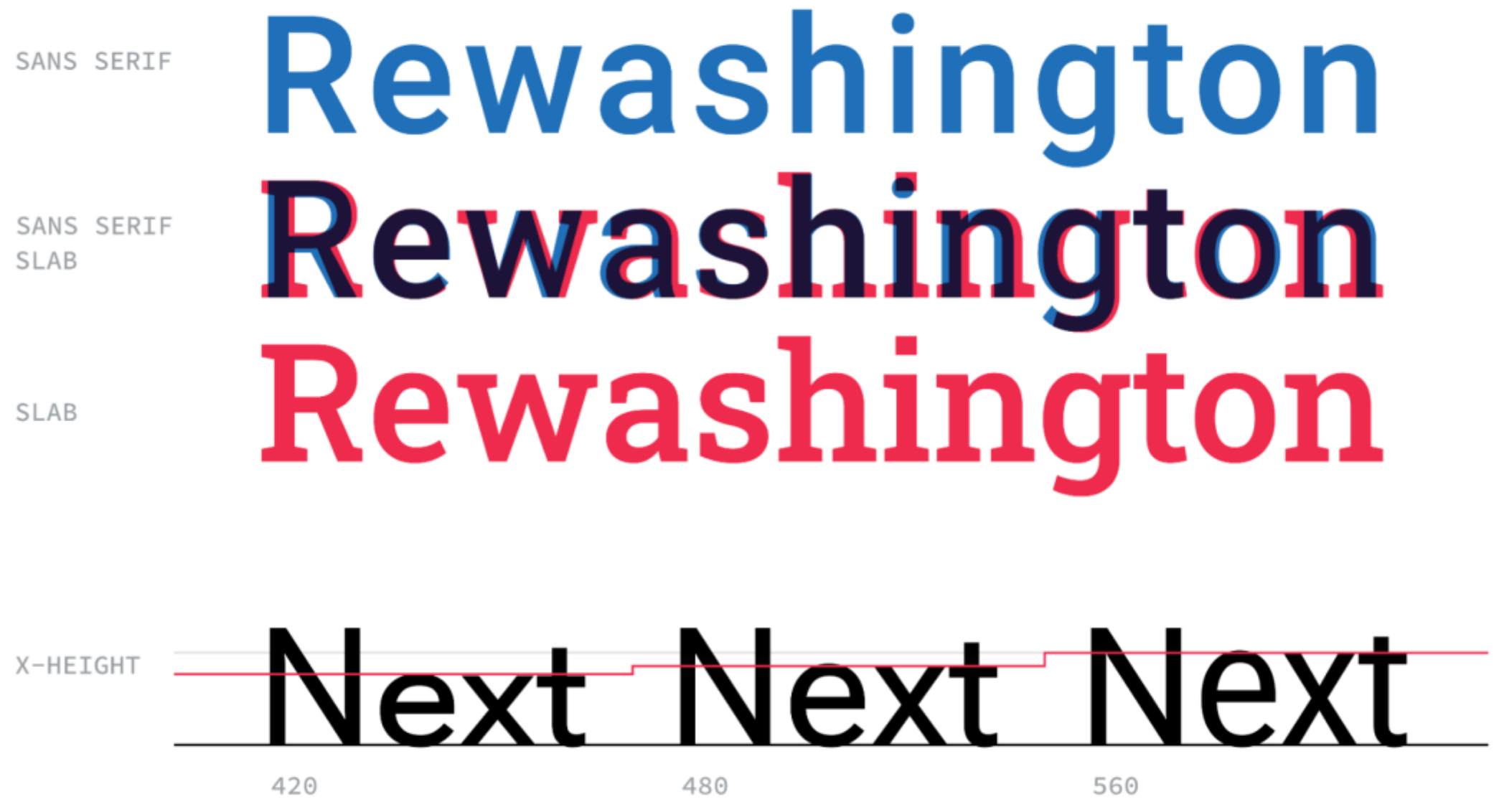

**Figure 7.** (top) Sans Serif (blue) and slab (red) fonts from the Roboto family. The two fonts are superimposed to visualize the differences in letter weight, width and serifs. (bottom) Variable font Roboto-Flex with x-height being adjusted to demonstrate the effect on the visual appearance of the font.

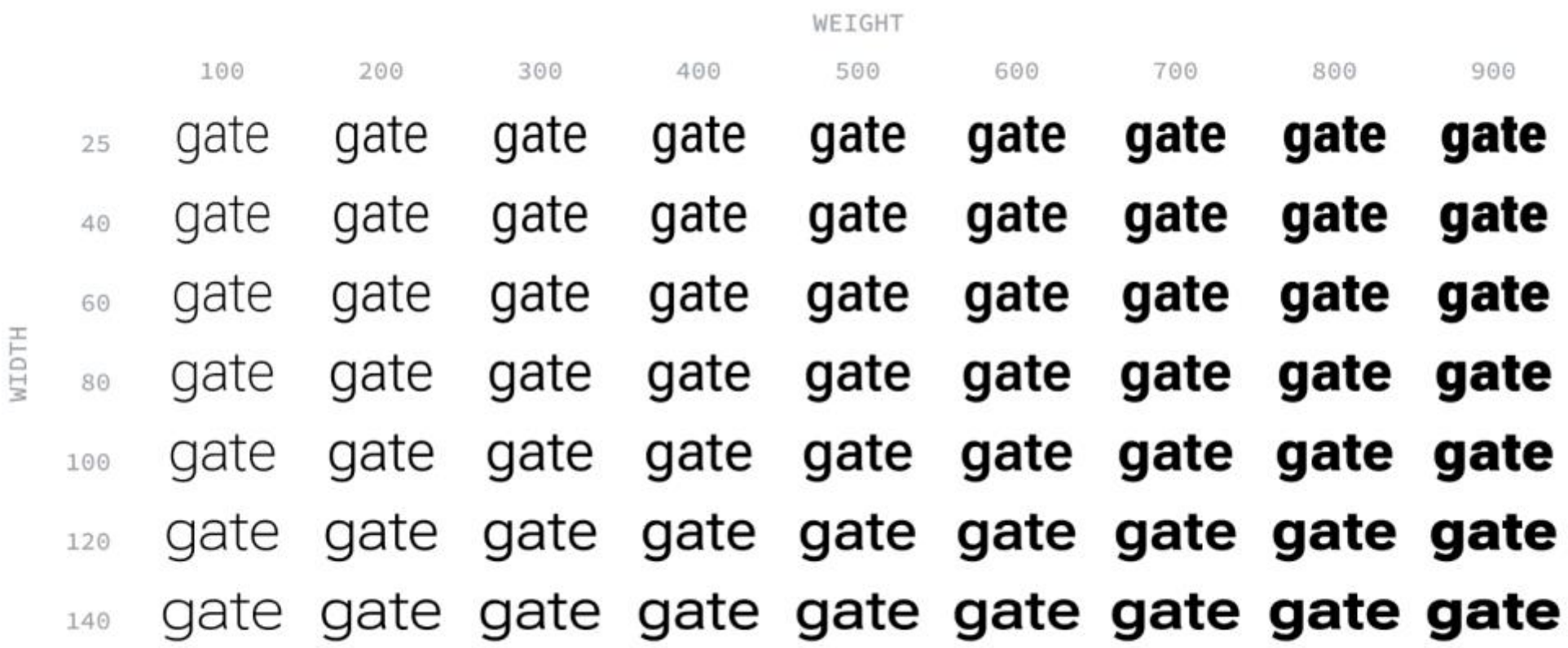

**Figure 8.** A typeface can include many variations of weights and widths, such as this example made with variations of Roboto-Flex, a variable font.

For example, the presence of serif on a font has been shown to lower reading speed at small sizes compared to the same font without serifs (Morris et al. 2002), yet in other reading situations, serif can improve recognition of single letters on vertical extremes at a distance (Beier & Dyson 2014). Low stroke contrast improves word recognition (Minakata et al. 2020). Simple letter skeletons result in greater letter recognition (Beier & Larson 2010; Beier et al. 2018). Condensed fonts impair letter recognition (Oderker & Beier 2020), and so do heavy and light letter weight fonts (Beier & Oderkerk 2019a), which also slow down reading speed (Chung & Bernard 2018).

**Perceptual size matters.** Traditionally, many studies have focused on comparing different typefaces such as Arial and Times New Roman and comparing these in the same fixed point size per condition (Bernard et al 2001; Wallace et al. 2020). This approach may introduce confounds, as the perceptual size of a font is dictated by its x-height (distance between the baseline and the mean line in a font) (Fig. 9), and not its point size (Beier 2012). In recent times this has led to efforts to present stimuli fonts at a perceived font size by comparing fonts set at similar x-height (Wallace et al. 2020 VSS; Xiong et al 2018; Beier & Oderkerk 2019b). Most fonts contain OS/2 tables that show various parameters set by the authors, such as x-height and length of ascenders and descenders. However, not all fonts have accurate OS/2 values.

**The problem of interacting variables.** To identify the effects of specific font properties that can be transferred to other fonts, research will need to isolate the significance of properties. This can be done by comparing fonts belonging to the same typeface family (e.g., width variation between Univers Condensed and Univers Expanded), or design fonts for the experiments where all other possible interfering variables are controlled for (Beier 2013; Gürtler & Mengelt 1985; Beier & Oderkerk, 2019a, Chung & Bernard, 2018).

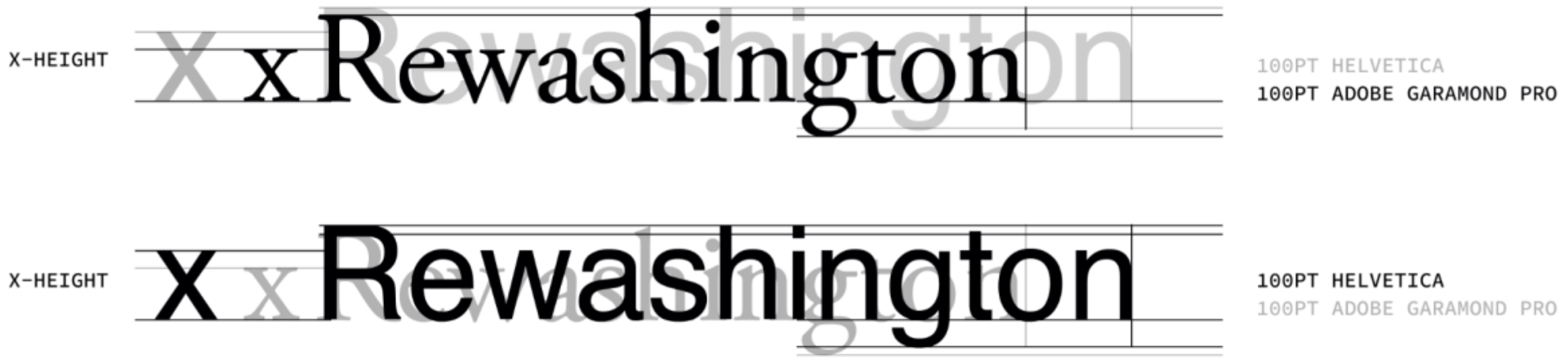

**Figure 9.** A comparison of two fonts of different x-height set at identical font sizes (Helvetica Regular and Adobe Garamond). The different x-heights result in Garamond having longer ascenders and descenders, as well as appearing to have greater leading between the lines of text and having a smaller font size.

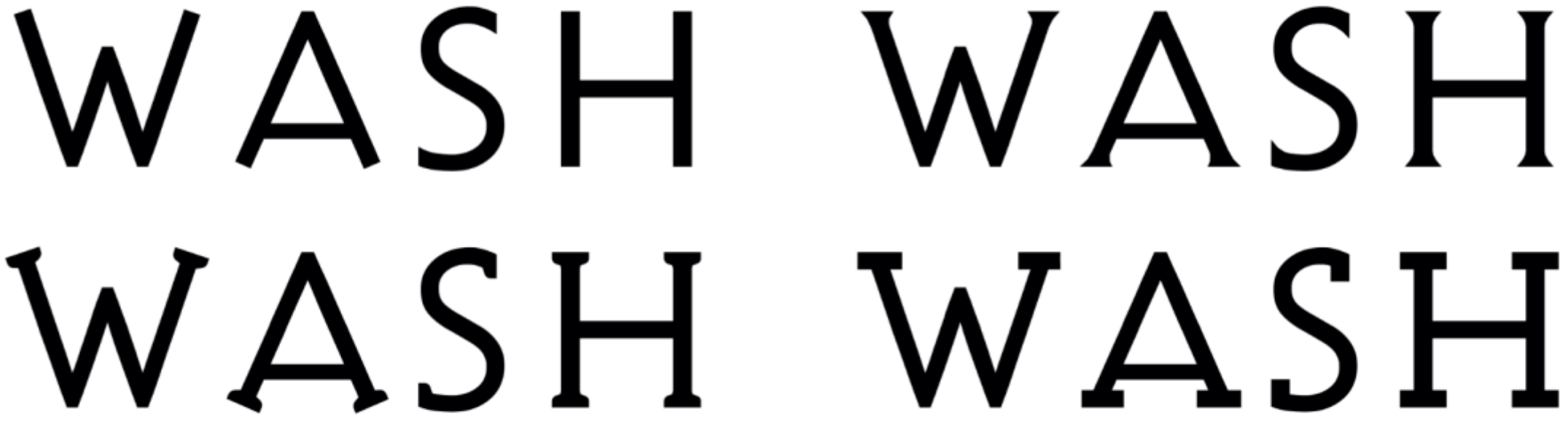

**Figure 10**. Decovar, a Google Font, designed by David Berlow, Font Bureau, 2017, demonstrates how different serif structures can co-exist in a single variable font file.

**Variable fonts.** The introduction of Opentype 1.8 in 2016 has made it possible to have many fonts in a single file (Figure 10); this is known as ***variable fonts*** (Hudson 2016). This format allows for the different properties of the font to vary on a single or multiple axes with extreme instances at the ends. For example, the weight of a bold font, the width of an expanded font, or the stroke contrast, are not predefined. The user can choose the exact coordinates on one or more axes. This flexibility can enable researchers to be more in control of the magnitude of each font variable (Figure 7,8).

## 4.2.4. Controlling Typographic Settings and Environments

**Control typographic settings.** In addition to controlling the font selection and properties, typographic settings should ideally be controlled. Some of the typographic settings that have shown to influence reading are letter spacing (Perea & Gomez 2012), word spacing (Slattery & Rayner 2013), contrast polarity (Dobres et al. 2017), background complexity (Sawyer et al 2020), and font color (Ko 2017). A common approach is to use cascading style sheets, or CSS, to manipulate the typographic setting (Wallace 2019). CSS contains font properties that can be used in the analysis of font weight and word spacing and can be further used to manipulate the page layout of the text presentation. For example, the background color and the width of the paragraph can be manipulated to affect visual span.

**Consider how fonts may be perceived.** In addition, familiarity with the text and its presentation is of importance. Unfamiliar fonts (Beier & Larson 2012; Zineddin 2003) or unfamiliar script styles (Pelli 2006; Ngiam et al 2018) can negatively influence reading. Also, given the apparent agreement on perceived font personalities (O'Donovan 2014), text stimuli need to be controlled for semantics, vocabulary and context (see also Section 4.1). Recent work in this space explores the space of matching recommendations (Shirani 2020, Kulahcioglu 2020), and bespoke fonts for a given context (Wang 2020).

**Consider what purpose different fonts were designed for.** It should not be assumed that all fonts are equally appropriate for testing on all reading platforms or for all reading situations. Many large-size typeface families include fonts of different optical sizes, where the fonts for smaller sizes typically have larger x-height, low stroke contrast, and greater width and spacing (Ahrens & Mugikura 2014). Many typefaces were designed and engineered for specific rendering systems (e.g., Microsofts' ClearType fonts (Berry 2004)). Typefaces can also be designed with specific attention to how they will be used by content developers (graphic designers, web designers, app developers, ux/ui developers), and indeed many other specific applications or users. Most fonts designed for use on-screen will work well in print, while not all fonts designed for print will work well on screen, and some fonts designed for large sizes will not work well in small sizes.

To ensure external validity, a central question is whether test stimuli presented on one type of screen and resolution automatically transfer to any other screen or printed matter. New reading formats are constantly emerging. It is difficult to predict if findings reached through the use of traditional screens can be transferred to reading on new displays such as AR and VR.
Mixing and matching different font formats with different digital representations may have a negative effect on the validity of experiments if rendering and resolution are not controlled for, it is very likely that a specific font or typographic variable will benefit reading in one environment but not another.

### 4.2.5. Considering Resolution and Rendering

Screen optimization of fonts usually includes: large x-height, open ***apertures***, large ***counterforms***, generous letter spacing, limited stroke contrast and ***delta hinting*** (Fig. 11). Other central variables to consider are rendering (Ahrens, 2021), font size, resolution, browser, and operating systems (Baset, 2020).

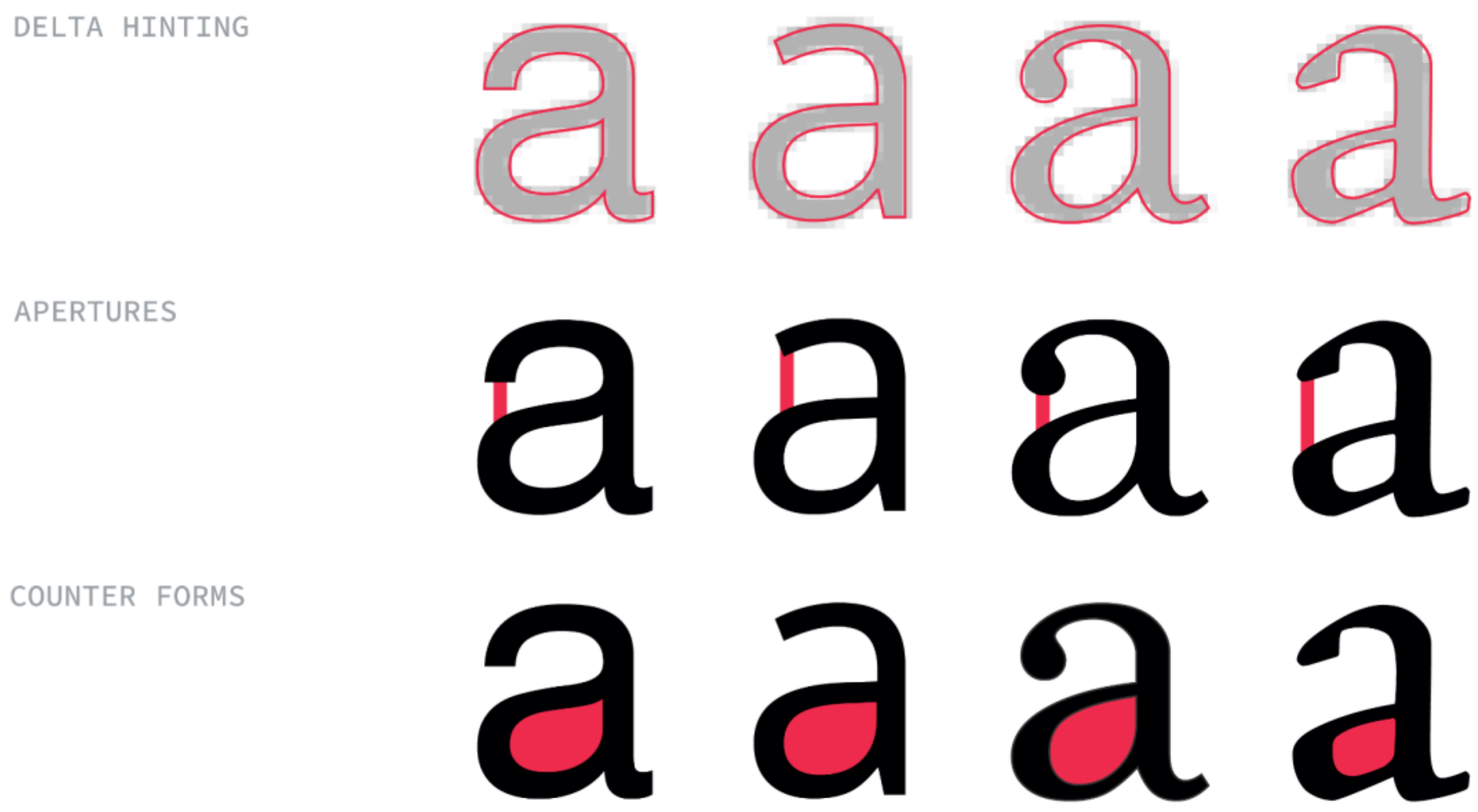

**Figure 11.** Delta hinting, apertures, and counter forms are three critical features of a font that can significantly affect their appearance on digital screens.

Nearly all fonts store their outlines as ***Bézier curves*** so they can scale to any size without losing fidelity. A notable exception is bitmap fonts, which use images instead of resolution-independent vectors. Bitmap fonts are often used to display emoji (though there are also vector-based emoji fonts) and special care must be taken when using these at large or small font sizes.

There are two competing approaches to storing Bézier curves in a font, TrueType by Apple and Microsoft, and the Compact Font Format (CFF) by Adobe. The primary difference between these is that TrueType uses quadratic bézier curves while CFF uses cubic bézier curves and the way they perform hinting. The difference between quadratic and cubic beziers is not a concern at common reading text sizes, but the difference in the hinting approach can have a significant impact on the legibility of a font, as explained in the following.

Software libraries called **rasterizers** turn vector fonts into pixels for display on the screen. Most rasterizers are part of the operating system, but there are also standalone rasterizers that can be used on multiple operating systems. There are four rasterizers in common use: GDI and DirectWrite on Windows, Core Graphics on macOS and iOS, and FreeType on Android, Linux, and ChromeOS. Rasterizers turn Bézier curves into pixels by sampling them at the desired resolution. At its most basic, the rasterizer checks whether each pixel is inside or outside of the curve. If the pixel is inside the curve it is colored black. If this sampling is done at a high enough resolution or a large enough font size, the result is a near-perfect approximation of the curve.

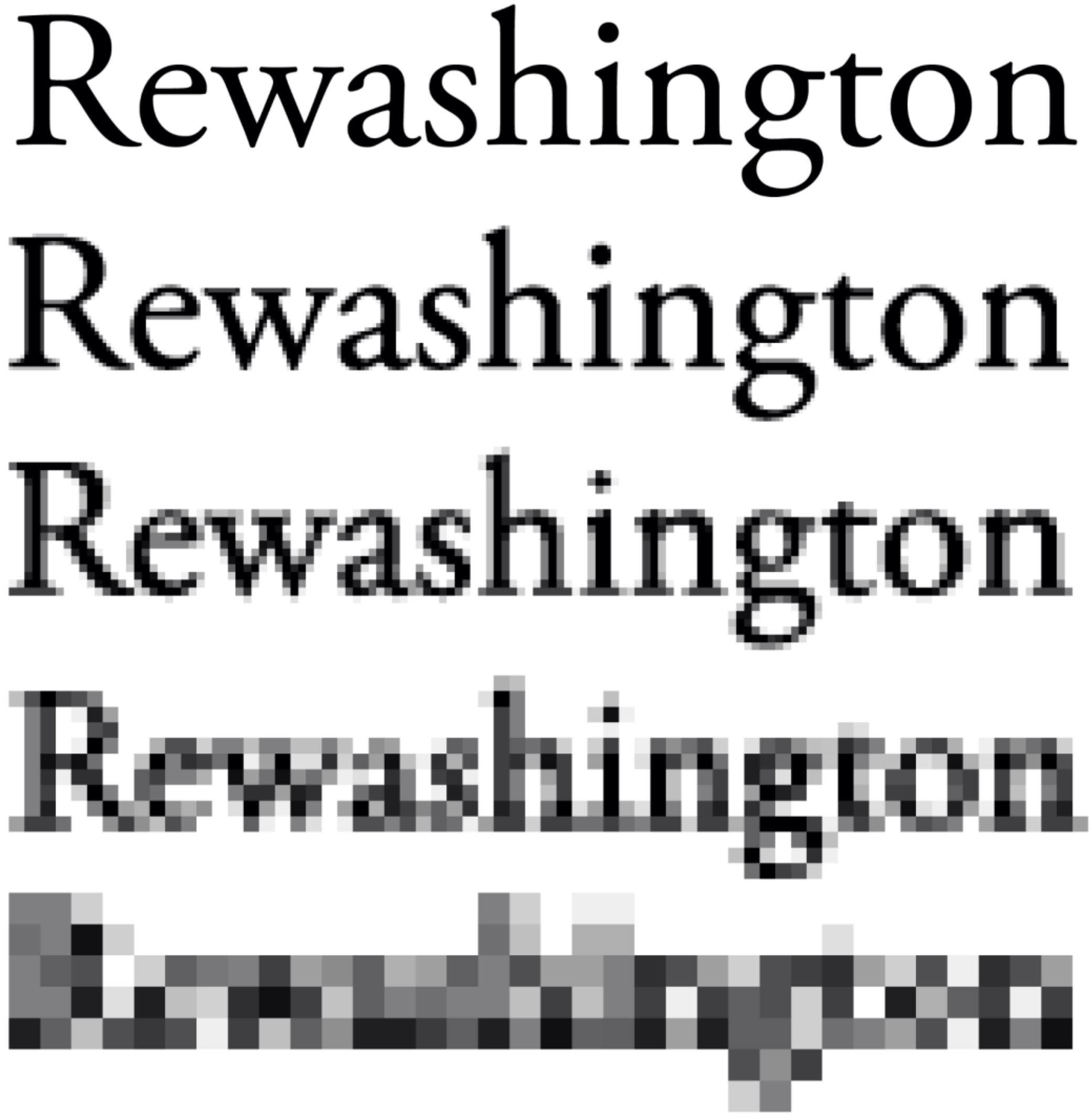

**Figure 12.** Rasterizing at different resolutions and font sizes.

Resolution and font size are thus linked. High-resolution screens produce good results at low and large font sizes, but the legibility suffers when small font sizes are used on a low-resolution screen (Fig. 12). To address these issues, fonts include hinting instructions, which tell the rasterizer how to behave at low resolutions. TrueType fonts include these hinting instructions in the font itself, while CFF-based fonts rely on the rasterizer to fix these issues. The TrueType approach gives more control to the type designer but is also more time-intensive and thus expensive. Typefaces that include extensive hinting are often advertised as especially geared towards legibility at small sizes. Not all rasterizers support hinting, for example, Apple's Core Graphics rasterizer ignores hints because Apple's devices generally have high-resolution screens which reduce the need for extensive hinting.

If possible, favor extensively delta hinted typefaces for legibility and readability studies. While some rasterizers may ignore the hints, others will benefit from having high-quality hints. It is believed that this simple rasterization approach works fine for printers because they have very high resolution. Unfortunately, screens don't have such a high resolution yet, so rasterizers perform antialiasing to approximate smooth curves at low resolutions. There are two types of antialiasing: grayscale and sub-pixel. Grayscale uses grayscale values to approximate partially filled pixels, while subpixel antialiasing uses a screen's red, green, and blue sub-pixels to achieve the same. While sub-pixel rendering may sound superior, it is often disabled because subpixel antialiased text cannot be rotated. The subpixels in a physical screen are fixed, so subpixel antialiasing only works well on screens that have a single, fixed, orientation. Sub-pixel antialiasing also produces noticeable color fringing on low-resolution screens which can be distracting to some readers. To make matters worse, sub-pixel antialiasing is implemented slightly differently in each rasterizer. Notably, there is presently little research to determine whether these processes lead to enhanced readability, and what little exists calls into question whether they are even detectable by human observers (see Hancock, Sawyer, and Stafford, 2015).

## 4.3. Licensing

To be able to freely use content in reading studies, the source material needs to have the appropriate use licenses attached to it. This is particularly relevant to industry partners who may use the results of reading studies to inform commercial applications and future product development. The U.S. Copyright Office Fair Use Index is an educational exemption for using copyrighted materials in the classroom (U.S. Copyright Office 2016). To reuse content for research within an academic setting, it is advised to consult with your university’s or organization’s legal counsel to determine if your research meets the standards of the fair use exemption. You might also look for text available in the Public Domain or under Creative Commons licenses.

Content creators can also be open to having their content, whether it is full texts, excerpts or fonts, used for research purposes. It is critical to work with the property owners to get permissions that allow for research while protecting their intellectual property and to establish a mutual understanding of how the results will be shared. Font's End User License Agreement (EULA) varies greatly in how the user is permitted to use the font and how much the user is permitted to alter the font. If you are interested in altering an existing font, it is suggested to seek out open source fonts where adjustments are permitted. In any case, if you are using fonts for research, it is recommended to ask the copyright holder. Copyrights can be found in the font data. When in doubt, researchers can inquire on typedrawers.com.

It is also strongly desirable that all results, whether supportive or not of the content used, are shared for the value such learnings add to the broader field. We recommend making all data publicly available, as it may affect past and future readability data analysis. Indeed, we submit that the lack of easily accessible materials, and of unfavorable results, is a significant impediment to research in this space, and we see this as an important area of future growth of the field. We urge educators, designers, researchers, and content creators to collaborate in assembling and freely disseminating properly-leveled reading materials (passages and support materials - e.g., comprehension questions) for different populations of readers, and we urge researchers to recognize the importance of the visual representation of texts and their possible impacts on study results.

# 5. Equipment, Devices, and Software Tools

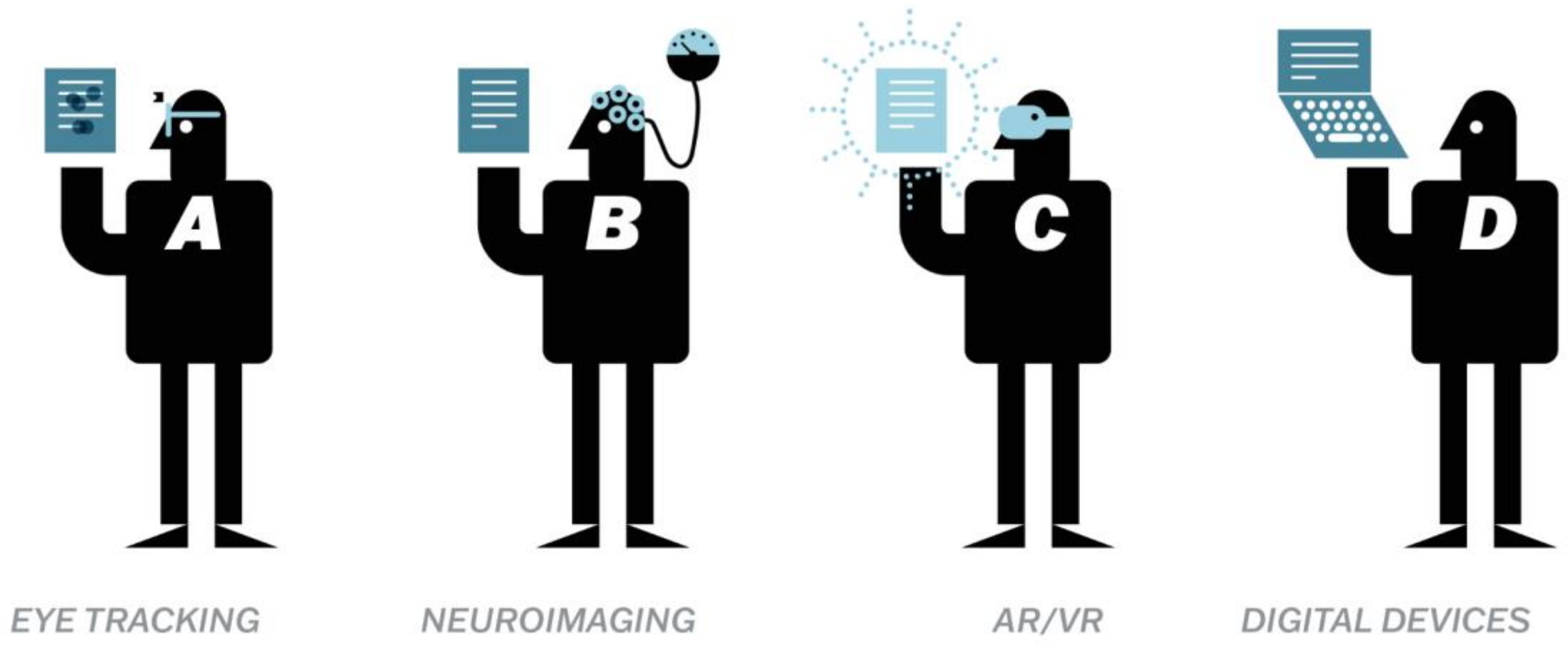

**Figure 13.** Fundamental tools for understanding readability have changed dramatically, with the promise of more change to come.

People read on many different platforms and in many different contexts, from glancing at road signs to scrutinizing news articles (MacFadyen, 2011; Margolin, et al. 2013). In this section we discuss experimental set-ups for studying how readers read on digital devices, from brain imaging and eye tracking devices in the lab, to virtual experiments designed for the web (Fig. 13). The appropriate hardware and software to employ for a reading study depend on the context and environment, the target reader populations, the specific research questions, and the availability of resources. This section offers some considerations, some recommendations, and some weighing of factors in favor of the use of different experimental equipment, hardware devices, and software tools.

## 5.1. General-Purpose Digital Displays

Modern reading has grown significantly more complex since the widespread adoption of the digital display in the 1970s. Where once text was read solely from printed material, reading is now done on a wide variety of display types: large desktop monitors, high resolution smartphones, lower resolution purpose-built displays (kiosks, in-vehicle dashboard displays, etc.), specialized e-ink devices, smart watches, etc. Early studies of the legibility of digital text suggested that it was inferior to traditional printed material (Mills & Weldon, 1987). More recent studies suggest that as the resolution and fidelity of displays has improved, displays have achieved parity with print in terms of pure legibility (Margolin et al 2013), though readers may be able to maintain better awareness of their performance with print (Clinton 2019).

**LCD displays versus e-ink displays.** One key difference between print and digital displays is that print (and e-ink) reflect light, while digital displays emit light. E-ink displays (e.g., like Amazon's Kindle) have no backlight. There has been a body of research comparing the two media. Results have been mixed, and in aggregate suggest that print/e-ink and digital LCD displays have equivalent practical legibility (Siegenthaler et al 2011; Siegenthaler 2012; Lee et al 2008). Differences in legibility between display types may in fact have more to do with the amount of illumination, both in the environment (Lee et al 2008; Dobres et al 2017) and in the amount of light being emitted directly from the screen (Dobres et al 2016). Research suggests that lower illumination settings cause the pupil to dilate over the imperfect surface of the eye, thus exacerbating the effects of astigmatism and smaller flaws in the lens, ultimately hindering legibility (Piepenbrock et al 2013; Taptagaporn & Saito 1989). These findings present a particular problem for increasingly popular “dark mode” designs, which are self-reported to reduce eye strain (Eisfeld & Kristallovich 2020), but may have reduced legibility due to their lower-light nature.

**Capturing reader behaviors on digital displays.** General-purpose digital displays (Yeykelis, Cummings, and Reeves, 2014) provide the potential to reveal how readers are moving through a piece of text through incidental movements. Such movements can be captured without impinging on natural reading behavior, a key consideration for understanding how readers read in real-world situations. These can include a reader's click-stream as they move through a document, where they position their mouse (Cooke, 2006; Huang et al., 2011), and how they scroll through a document (Fitchet & Cockburn, 2009), including multi-touch behaviors on phones and tablets. Screenshot software can be employed to determine what participants have on screen (Brinberg et al. 2021; Reeves et al. 2019) but can also include gyroscope data from modern smartphones (Pires et al., 2018) which can reveal, for example, whether a reader is engaging with a text while walking through the orientation of their device (Barnard et al., 2007; Mustonen et al., 2004). Finally, audio recordings, which can be supported by any device with audio input, can be used to approximate reading activity through read-aloud protocols (Banerjee et al. 2011; Bernard et al., 2001; Rello et al., 2016).

## 5.2. Research Equipment

### 5.2.1. Eye Tracking

Since reading requires a reader to move their eyes from word to word along a line of text (e.g., to make saccades from one word to the next), ***eye tracking*** (Table 1) has been a key technique in reading research since it was first developed more than a century ago (Javal, 1879; Rayner,1998; Tinker, 1946). Tracking where a reader looks whilst they read can reveal what words in a sentence they skip, whether they backtrack, and how they move through a passage--and, potentially, show what visual strategies they are adopting based on the type of reading (see Section 2.1). That being said, while eye tracking is a powerful method, there are significant limits on what gaze information can tell researchers, since fixating at a word is no guarantee it was read or understood (Drew et al., 2013), and deducing what a reader was doing based only on where they looked is difficult since it requires knowing how to classify different types of reading based on gaze behavior, what the reader's task was, and whether that task is appropriate to the text they were reading.

**Hardware-based eye tracking.** A range of eye-tracking equipment exists, typically in the form of non-intrusive hardware that uses near-infrared light to create reflections on the eye and, in conjunction with a camera pointed at the participant's eyes, uses these reflections to infer the eye position, orientation, and its movements (Hammoud, 2008). This specialized equipment comes in two main forms: head-mounted systems (Cognolato et al., 2018) and remote systems (Niehorster et al. 2018). Head-mounted systems structurally resemble eye glasses and are preferred in naturalistic studies that involve a lot of movement, for example, in marketing research on product placement (Hendrickson & Ailawadi, 2014). Remote systems are stationary and the eye tracker is mounted near or integrated in a monitor. These eye trackers are capable of higher performance and accuracy compared to head-mounted systems. These characteristics can be further augmented when the eye tracker is coupled with head stabilization (e.g., using chin rests or bite bars) which keep the participant at a constant distance from the screen and minimize movement. Additionally, remote eye trackers are often more appropriate for reading studies as they are mounted on the specific screen that the reading task will take place on. In the case of reading studies on mobile devices, remote, standalone eye trackers must be mounted along with the mobile device on a specialized stand. On the other hand, head-mounted systems allow for movement and can be used in non-digital settings but may create challenges on mapping the gaze on specific areas of interest due to the distortion of the scene that comes with movement. Beyond desktops, laptops, and mobile devices, eye tracking have also been embedded in virtual and augmented reality devices. Some popular eye tracking manufacturers include SR-Research, Pupil Labs, and Tobii which offer a range of research- and consumer-grade systems.

**Camera-based eye tracking.** Recently, software solutions that use standard webcams have been developed as an answer to the high cost of such equipment that can rise to tens of thousands of dollars with a compromise in the quality of data acquired. They are browser-based (e.g., Papoutsaki et al. 2016), desktop (e.g., Zhang et al. 2019), or mobile applications (e.g.,

Krafka et al. 2016). While the research-grade systems can offer extra capabilities, such as pupillometry to measure cognitive load, or alertness, in addition to superior performance and accuracy, the democratization of eye tracking could enable psychophysiological research to be conducted on a large scale in many different environments.

**Table 1.** Advantages and Disadvantages of Eye-tracking System Types

| Eye Tracking System | Typical Environments | Advantages | Disadvantages |
|---|---|---|---|
| **Head-mounted** | Out of the lab studies which can include a lot of movement (e.g., sports, driving, marketing). | Allow for free movement of the participant. | Lower accuracy, precision, and volume of data compared to remote eye tracking. More difficult to map the gaze locations on text and perform statistical analysis. |
| **Remote** | Lab studies that aim for highly-controlled experiments. | High volume of data compared to head-mounted eye tracking. Data is directly mapped to the screen containing the reading task which simplifies subsequent analysis. | Little to no movement is allowed to acquire data which can lead to unnaturalistic behavior. |
| **Web Cam** | Experiments deployed with remote crowdworkers. | Can be run on many participants, in parallel, to produce a large volume of data,from potentially diverse participants in their naturalistic environments. | Lack of control over data quality, noise, and confounding environmental factors. |

## 5.2.2. Neuroimaging

Neuroimaging research can inform our understanding of which brain regions and networks are active during reading, as well as the underlying processes of reading. Since reading is a visuo-cognitive process, non-invasive neuroimaging techniques like electroencephalography (EEG) and functional Magnetic Resonance Imaging (fMRI) have the potential to reveal internal cognitive and linguistic processes that are otherwise inaccessible to researchers.

**EEG Systems.** Electroencephalography is used to measure electrical activity in the brain using non-invasive electrodes placed on participants' scalp while the participant is performing a cognitive or linguistic task of interest. Choosing an appropriate EEG system depends on the population being studied and the goals of the experiment. For instance, the number of electrodes in an EEG system varies greatly from just a few electrodes up to 256. Systems with more electrodes will naturally require a longer and more extensive setup, but will provide better localization, that is, where activity is occurring in the brain. However, a long setup time may not be well tolerated by some populations, such as children. Another distinction can be made between wet and dry electrodes. Wet electrodes are so named due to the electrolytic gel that needs to be applied to the scalp to serve as a conductor. Dry electrodes make setup much easier, but the signal to noise ratio is worse. Finally, some experiments may benefit from using a mobile EEG system, which allows participants greater freedom of movement when compared to a traditional EEG system, at the cost of a coarser-grained and noisier signal. A challenge when using EEG for reading studies is the noise introduced by eye movements. Methods such as independent component analysis (ICA) allow for researchers to identify and remove eye blinks from the signal (Jung et al., 2000). Additionally, some researchers are combining EEG and ICA with eye tracking to better identify the relevant signal (Dimigen et al., 2011; Plöch, Ossandón & König 2012).

**ERP components.** Event-related brain potentials (ERPs) are waveforms extracted from EEG, and are believed to be generated from the summed activity of specific cortical neurons (Peterson, Schroeder, & Arezzo, 1995). ERPs have excellent temporal resolution and are therefore prime candidates for investigating the time course of multiple rapid processes that underlie reading comprehension. Specific ERP components are indicative of different types of processing. The names of ERP components often begin with an "N" indicating it is a negative-going component, or a "P" indicating it is positive-going. Some ERP components frequently examined in reading research include the N250, which likely reflects form-based processing (Holcomb & Grainger, 2007), the N400 which reflects semantic processing (Kutas & Federmeier, 2011; Kutas & Hillyard, 1980), and the P600 which reflects syntactic processing (Osterhout & Holcomb, 1992). Researchers can compare the effect of various manipulations on the latency, amplitude, and scalp distribution of the ERPs of interest. They can also compare ERPs across different populations.

An exciting new development in EEG/ERP readability research is the creation of reading corpora associated with EEG data (Frank et al., 2013; Hollenstein et al., 2018; 2020). For example, the Zurich Cognitive Language Processing Corpus (ZuCo 2.0) has made publicly available EEG data of participants completing a natural reading task, where they read at their own pace, compared to a task-based reading task (Hollenstein et al., 2018; 2020). Simultaneous eye-tracking data were also acquired and are available in the corpus. We are hopeful that researchers leveraging EEG/ERP techniques, or other neuroimaging, in their own readability research will be open to releasing additional sets relating a reading corpus to brain activity recordings.

**MRI and fMRI.** Structural Magnetic Resonance Imaging (MRI) generates high resolution images of the brain, with the ability to distinguish between different types of tissues (e.g., gray matter and white matter) and brain structures. In addition to acquiring structural information, MRI can be used to investigate functional activity in the brain using methods like functional Magnetic Resonance Imaging (fMRI). When a particular area of the brain is engaged by a task of interest the blood becomes more oxygenated in that region as neural activity increases. fMRI is sensitive to the blood oxygen level dependent (BOLD) signal as a marker of brain regions that are more activated during a task of interest.

fMRI has been used to investigate the processes underlying reading comprehension within specific brain regions, such as the visual word form area (VWFA) which is important for decoding written words (Cohen et al., 2002; Dehaene & Cohen, 2011). Researchers can also investigate brain networks involved in tasks like reading. This is accomplished by examining functional connectivity, which is defined as the coactivation of multiple brain regions during a task of interest. Investigating functional connectivity can help us understand the overall organization of reading in the brain. For example, research suggests that children with developmental dyslexia have disrupted functional connectivity between left occipitotemporal, left inferior frontal, and left inferior parietal regions that are important for reading comprehension (van der Mark et al., 2011).

fMRI can also be used to investigate the contribution of specific regions and brain networks in specific populations of interest such as young developing readers, or dual language learners. For instance, Gaillard and colleagues (2003) found that the reading network in young developing readers is very similar to the reading network in adults. As another example, Meschyan & Hernandez (2006) investigated the neural networks activated during reading in Spanish versus English in a group of Spanish-English bilinguals. They found that the less proficient language activated the brain's articulatory motor system, and was associated with slower reading times.

**Comparing EEG and MRI.** Electrophysiological methods like EEG have excellent temporal resolution, which is an advantage when studying reading, where many processes occur in quick succession. However, EEG does not have sufficient spatial resolution which means the specific brain areas involved in various processing steps cannot be inferred from EEG alone. Researchers have been developing high-density EEG systems combined with source localization algorithms to enhance the spatial resolution of EEG (Michel & Brunet, 2019). Methods like fMRI have relatively poor temporal resolution, but they do have excellent spatial resolution. Techniques like fMRI are used when researchers are interested in investigating the specific brain regions activated during reading tasks. Recent work in cognitive neuroscience shows promise for "fusing" the temporal resolution of EEG with the spatial resolution of fMRI via analysis techniques, for a deeper understanding of brain processes (Cichy & Oliva, 2020). Unfortunately, high quality neuroimaging systems and data analysis are quite expensive and require specialized training to use, calling out the need to collaborate with neuroimaging specialists for studies that require use of these methods.

### 5.2.3. Virtual Reality and Augmented Reality

Virtual Reality (VR) and Augmented Reality (AR)–often referred to as Mixed Reality (XR)–have gained popularity as a research platform. VR simulations are often used for training and educational purposes as users adopt similar behaviors in VR as they do in the real world. Studies in VR allow researchers to put participants in different real-world scenarios and investigate different in-field environments at scale (Mäkelä, 2020). Platforms such as the Hololens or Vive Pro headsets (Microsoft, 2021; Vive, 2020) have higher fidelity head and eye tracking than mobile devices as well as allow for 3D interaction with the content. Such platforms provide opportunities and new challenges for readability research.

**Text for augmentation.** AR applications are designed to enrich users' physical activities with digital information overlaid visually. This paradigm makes digital text powerful by tying in the context of on-going activities—for example, remote assistance and responsive instructions (Wisotzky et al, 2019) used in industrial training or interactive gaming experiences (Kim et al., 2019; Ružický et al., 2020; *Pokémon GO* 2021). A common challenge is that text reading happens as users engage in other activities in parallel. It can be difficult to ensure users see the text when their mental load is high, (Lindlbauer et al., 2019) and continuously changing surroundings as background textures can cause legibility issues (Gabbard et al. 2006). Therefore, parameters such as textures in the AR background, lighting, user's attention-levels, and mental load can contribute to the overall text readability.

**Challenges for rendering text in AR/VR.** Reading in mixed-reality environments is becoming more prevalent with the commercialization of consumer devices and advances in display technology that allows high-quality text renderings. When e-books were introduced, a large body of research focused on comparing paper to digital screen reading. Mixed reality reading studies have not received that same attention yet. These platforms introduce readability challenges when presenting text in simulated 3D environments or when superimposed over the ambient environment in AR settings. The placement of text with AR can be a safety consideration, and early work showed that users preferred that it was placed in uniform regions (Orlosky et al., 2013). Rzayev and colleagues investigated reading on head-up displays and the effect of text position and presentation type on the reading experience (Rzayev, 2018). Effective positioning depends on the user’s primary task: when focused on the reading, a center position in an AR display allows for best comprehension. During walking, on the other hand, shifting the text position to the bottom center helps users to keep track of their path while reading. In immersive displays, resolution can limit readability and research has shown magnification and augmented floating text lead to favorable experiences for users (Knaack et al., 2019). Other research has explored optimal readability settings for font and distance (Büttner et al., 2020), as well as text size and positioning (Dingler et al., 2018).

**Opportunities for immersive reading.** Other works have started to explore text renderings on 3D objects where text is warped across concave or convex surfaces (Wei, 2020) and text interaction in virtual environments (Dingler, 2020). Virtual environments have the potential to immerse the reader in multimodal reading experiences where the visual, audio, and haptic environment adjusts to the content. We expect to see much more research in this regard with novel reading applications that immerse readers and provide innovative reading experiences. Notably, while some VR headsets come with eye-tracking capabilities, they may not come with the licenses that allow for experimenters to actually use the data; we advise caution and careful reading of licenses in such equipment.

To work out readability parameters for mixed reality text renderings, such as which font family is most adequate in 3D environments, lab studies offer a controlled way to guide participants through different reading conditions. Compared with studies on 2D displays, participants may be subject to motion sickness and increased visual fatigue. Experimenters should allow participants to take frequent breaks and monitor motion sickness symptoms.

**Taking viewing distance into account.** In VR, readers are fully immersed into an environment as their visual, auditory, and even haptic and olfactory channel (Brooks, 2020) can be catered to. For readability, text display parameters, such as the angle size are important design considerations in VR. Google introduced a unit for perceived size in VR called 'distance-independent millimeter' (dmm), where 1dmm equals 1mm height at a 1m viewing distance. The unit allows to design layouts that can be applied to any screen at any distance. The perceived distance of objects in VR is dependent on vergence, i.e., the eyes' simultaneous pupil movements towards or away from one another when focusing. We explored font size, vergence, and view box dimension for comfortable reading in VR (Dingler, 2018) finding the median most comfortable distance to be 2.7m.

## 5.3. Software Tools

With a focus on readability, study designs in this space rely on an ability to experiment with the presentation of the underlying document text (i.e., typographical and format changes) to evaluate the resulting effects on the ease with which a reader can decode the document. Here we discuss platforms that allow for manipulating text formats for such purposes, including existing commercial tools and new research platforms.

### 5.3.1. Commercial Tools

Reading on digital devices, whether those devices are laptops, smartphones, tablets, or dedicated e-readers, brings with it a new set of considerations that have helped to spur readability research in the recent past. Readers using these devices, whether they are reading a PDF in Adobe Acrobat Reader, a webpage in a browser window, or an e-book on a dedicated device, are increasingly being offered opportunities to change the font, text size, character and line spacing, background color, and more to suit their individual needs and preferences. Adobe Acrobat Reader with Liquid Mode, Amazon's Kindle app, Apple iBooks, and Microsoft Immersive Reader are examples of reading applications with a subset of these readability features included. At the time of this writing, a current list of features in reading applications is summarized in Table 2.

**Table 2.** Typographical Manipulations Available by Reading Application

| Reading Application | | Character | | | | | | | | | |
|---|---|---|---|---|---|---|---|---|---|---|---|
| | | Fonts | Size | Spacing | Weight | Width | Line Spacing | Color | Justify | Page Width/ Columns | Notes* |
| Adobe Acrobat Reader | 1 | | ✓ | ✓ | | | ✓ | | | | ✓ |
| Amazon Kindle App | 2 | ✓ | ✓ | | | | ✓ | ✓ | ✓ | ✓ | ✓ |
| Apple Books | | ✓ | ✓ | | | | | ✓ | | | ✓ |
| Google Play Books Reader | 3 | ✓ | ✓ | | | | ✓ | ✓ | ✓ | | ✓ |
| Microsoft Immersive Reader Office/OneNote | 4 | ✓ | ✓ | ✓ | | | ✓ | | | | ✓ |
| **Browser Reading** | | | | | | | | | | | |
| Apple Reader Mode, Safari | | ✓ | ✓ | | | | | ✓ | | | |
| Google Play Books | | ✓ | ✓ | | | | ✓ | | ✓ | ✓ | |
| Google Reader Mode | 5 | ✓ | ✓ | | | | | ✓ | | | |
| Microsoft Immersive Reader Extension | 6 | ✓ | ✓ | ✓ | | | ✓ | | | | ✓ |
| Mozilla Firefox Reader View | 7 | ✓ | ✓ | | | | ✓ | | ✓ | ✓ | |
| Overdrive Reader | 8 | ✓ | ✓ | | ✓ | | ✓ | ✓ | ✓ | ✓ | |
| Reader Mode | 9 | ✓ | ✓ | ✓ | | | ✓ | | | | ✓ |

Instead of being limited by existing tools, the avid researcher can create customized reading materials by varying font features (like type, size, character and line spacing). Options include using Microsoft's Office Suite or similar document editors, working with variable fonts on support platforms (https://v-fonts.com/support), and using design software like InDesign.

### 5.3.2. Research Platforms

The Virtual Readability Lab (VRL) is a new platform containing several essential building blocks to engage users interested in self-paced studies. The VRL contains smaller 5-minute versions of previously published tests on reading speed and font preference (Wallace et al. 2020). It also contains additional 5-minute tests for users to find their optimal character spacing. The VRL allows other researchers to develop additional tests by using a unified database and building on current and future modules that passively track human behavior to enable studying reading behavior in-the-wild. The VRL also contains functionality to allow for teachers to sign-up their students and download their progress as each student takes various tests on the VRL to find out which font optimizations can improve their reading experience. The VRL relies on the voluntary participation of users by providing them insights about different ways to improve their reading behaviors, and it allows for users to compare themselves to the general population. This idea of motivating voluntary participation by providing to participants insights, scientific knowledge, and social comparisons has been successfully implemented by LabintheWild (Reinecke et al. 2015). LabintheWild has also conducted reading studies in the past, for example, exploring how webpage design affects reading experience and information retention, or reading and understanding different types of graphs.

Readability Matters has developed and made available the open-source Readability Sandbox. The Sandbox uses variable fonts to allow users to explore standard readability features such as font, font size, character spacing, character width, font-weight, line spacing, column width, and background color. Researchers can leverage this code for their testing purposes.

# 6. Experimental Methodologies

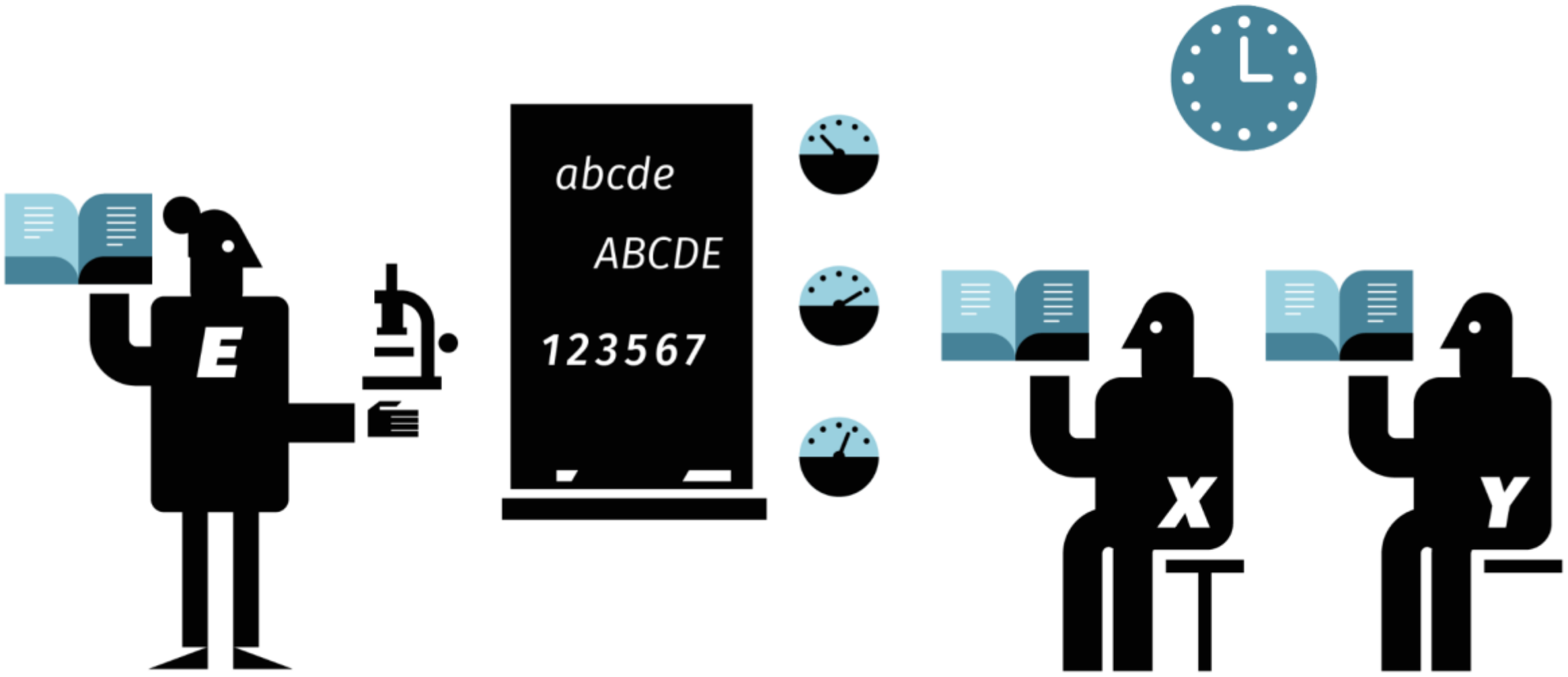

**Figure 14.** Readability literature contains great diversity in terms of experimental metrics and methods.

This section provides methodological guidance, with special attention to metrics, methods, and factors and considerations. The following is based on existing literature and the experience of the authors, and while not exhaustive, represents our perception of "core" readability methodology. The readability literature presently contains great diversity in terms of experimental metrics, methods, and factors considered, and as such what we present should be considered a foundation on which to build (Fig. 14).

## 6.1. Metrics

Readability interventions must be measured through metrics which gauge reader efficacy. Efficiency metrics should not be confused with "readability formulas" (see Crossley, Allen, and McNamara, 2011), predictive tools which exist to predict content readability before reading, relative to the level of the reader. Such predictive metrics are helpful in attaching material to a grade level, but are inherently unable to relate experimental manipulations to changes in reading efficacy. The metrics we present here all focus on efficacy, and in various ways measure readability by measuring factors which indicate success of the reader.

### 6.1.1 Reading Speed

Reading speed, often described as words per minute, is one standard metric for readability assessment. The speed of reading is often calculated in “words per minute” (WPM), calculated as the number of words read divided by the number of minutes taken to read them (Sawyer, 2003). Different researchers have found different ways to identify “words”, including absolute word count, number of characters per “standard word”, other aggregate numbers across paragraphs and pages, and still more schemes intended to smooth out the relative differences between written passages based on the stochastic nature of the language used. For example, a simple thought experiment will reveal that German prose likely involves more characters per a “real word” in English prose, and Chinese characters certainly provide possibly insurmountable challenges to any universality of the measure of “WPM”.

Mechanically, the speed of reading is a function of moving eyes across the page in a series of jerking movements termed “saccades“ and longer motions from the end of each line to the beginning of the next termed “return sweeps“. As a reader makes saccades, the distance of between stopping points is referred to as “ reading span“, and a larger span is unsurprisingly associated with faster reading. The speed with which readers perform return sweeps also varies, and in fact readers can be trained to make return sweeps more quickly in return for modest gains in reading speed. A simple thought experiment suggests that when the information on the page is not received by the reader, the reader must either forgo that information or read again. We have all experienced stopping mid-page to discover we have no memory of the content, even though our eyes have mechanically swept across every line (Sawyer, 2003). Indeed, the tendency to focus upon reading speed first and foremost obscures the fact that speed in reading is necessarily joined by the concept of accuracy in acquiring information, or reading comprehension.

### 6.1.2 Reading Comprehension

Reading comprehension is another standard metric for readability assessment. Comprehension in reading must be measured by probing participants' understanding of what they have read. The most common method for assessing reading comprehension is the use of “comprehension questions” relating to the material which has been read. Most studies, presumably for reasons of practicality, favor questions delivered shortly after reading, although naturalistic scenarios would seem to favor assessing comprehension further from the time of reading. Computing comprehension often takes the form of a percentage, where the number of correct comprehension questions are divided by the total number of comprehension questions.

The present lack of common and consistently used collections of passages and questions in readability research means that comparing between studies can be challenging as it is difficult to know whether the measurement instruments are comparable. Reading comprehension, therefore, is presently a standard metric which lacks a standard instrument of measurement.

Comprehension questions can be designed to tap a variety of comprehension strategies. Recall questions require readers to directly recall specific information from the text. Inference questions require readers to connect the information to fully understand the text. For example, questions can probe within-text inferencing abilities by requiring readers to connect information from multiple parts of the text. Summarizing questions require readers to combine main ideas presented in the text. Questions regarding the main idea or purpose of the passage can rely primarily on recall and recognition or, in more complicated texts, will require readers to synthesize information across the text and infer the main idea(s). Readers with strong inferencing skills are better able to fully conceptualize the text's purpose and meaning. For example, "Was Heraclitus of Ephesus renowned for a) philosophy b) textiles c) skateboarding or d) painting?" Practiced students can guess, and have honed the skill of answering formulaic comprehension questions even in the absence of knowledge. A related concern is the background knowledge of the participant, which can be addressed through surveys asking for reader familiarity or by removing questions that pilot participants indicate can be answered without reading the passage (Johnston 1984).

### 6.1.3 Speed and Comprehension Together

Speed in reading is joined by accuracy in acquiring information and understanding, or comprehension. The upper bound of speed at which readers can move their eyes from word to word will certainly have a negative impact on the ability of that reader to comprehend the material. Therefore, we can say that the aggregate effectiveness of a reader depends upon a speed – comprehension trade-off, likely with some similarities to classic speed-accuracy tradeoffs (Reed, 1973; McElree, Murphy & Ochoa, 2006). A speed – comprehension trade-off in naturalistic reading is not presently well understood and appears not to be a simple exchange but one contextually sensitive to, at the very least, reading purpose, material, and reader skill (see Section 2.1). A less skilled reader may move more slowly through a passage than a skilled reader, making reading speed a useful, simple measure of a reader's ability to move through a given passage. However, comparing reading speed between participants across passages can be difficult, since the participant and the passage may both be sources of variability in the data. To complicate matters further, slower reading can signal deeper engagement by a skilled reader. A participant's adjustments to their reading speed to compensate for the difficulty of the comprehension questions also represents this speed - comprehension tradeoff. When designing a study to measure reading speed, it is essential to counterbalance the order of passages and participants so that all passages are read the same number of times by all participants, and if the study involves varying typographical settings of stimuli, counterbalance this as well. That way, any possible difference found will relate to the difference of stimuli and not differences in the passages. We also recommend investigating the speed – comprehension trade-off in naturalistic reading in future research.

Fundamentally, reading comprehension is a more complex and subjective measurement than that of speed, and indeed reading comprehension is a more complex construct, requiring consideration of reader ability to retrieve, use, and integrate the phonological, morphosynthetic, semantic, and orthographic aspects of reading, as well as consideration of individual ability to recall background knowledge and synthesize this with the text (Alexander, Kulikowich, & Schulze, 1994; Elbro & Buch-Iversen, 2013). Matters of interpretation reveal that readers may even be required to understand the mental states of the author or the characters the author imagines. All of this complexity is filtered through necessary consideration of the cognitive processes involved in reading, starting with the transformation of the visual information presented upon a page or screen, passing through poorly understood intermediary processes, and ending in equally poorly understood mental representations. The challenge in clearly elucidating the measurement of reading comprehension, therefore its role in any trade-off, is clearly not small. Happily, great strides in understanding are being made in fields including neuroscience, vision science, cognitive psychology, and human factors.

## 6.1.4 Oral Fluency

For younger students, oral reading, i.e., reading aloud, is a standard process used by elementary school teachers to assess reading behaviors (Fuchs et al., 2001). Measurement tools, such as Running Record and QRI, evaluate oral reading fluency, including speed, accuracy, and prosody. Oral reading fluency, the speed at which accurate reading occurs, is expressed in Words Correct per Minute (WCPM), the number of words spoken correctly relative to their written form divided by the number of minutes the reading covers (Williams, Skinner & Floyd, 2011). Prosody is a more subjective measurement of expressive reading that measures appropriate correct timing, phrasing, emphasis, and intonation (Isardi, 1992).

Reading aloud can also provide valuable metrics for work with teen and adult readers. Oral reading can reveal reading format sensitivities with older students (Rasinski et al., 2017; Rasinski et al., 2005; Fuchs et al., 2001). Difficulties in reading aloud can be a result of difficulties in the visuo cognitive linkages necessary for fluent reading. Comparing this with readers reading silently can reveal where gaps exist in a given reader’s skillset, hinting at the process overall.

## 6.1.5 Phrase, Word, and Letter Identification

Some research focuses upon very short phrases, single words, and indeed individual letters. While reading at-a-glance is something naturalistically performed on electronic devices, some of these tasks have no applied equivalent. These methods instead probe sentence- and word-level processing, allowing researchers to carefully control stimuli in terms of the words themselves, varying factors such as word length, age of acquisition, or number of syllables. These methods also lend themselves to manipulations involving the presentation of each word with regards to orthographic and typographical characteristics of a text, as well as syntactic structure in the case of sentences. In word-level semantic categorization tasks, participants are asked to view single words and make a semantic decision about each word (e.g., “is it alive or not?”). Lexical decision tasks may also be used (i.e., “is this a real word?”), but semantic categorization tasks ensure participants comprehend stimuli to successfully complete the task. At the sentence level researchers often study sentence structure to see how syntax affects comprehension (Brothers et al., 2016; Brown et al., 2012, Sorenson Duncan et al., 2020).

Comprehension of individual letters and words via orthographic processing is complex, and must be understood in concert with integrating that information with syntactic and contextual information. Scientists have debated whether letter identification occurs primarily via a template-matching versus a feature-based paradigm, but most researchers now support a feature-based approach (Grainger, Rey, & Dufau, 2008). Thus, letter identification occurs primarily through the identification of individual features, such as horizontal lines, curves (e.g., open up vs. open down), and terminations. The set of features that are most important differ depending on the specific letter (Fiset et al., 2009). One measure of letter identification involves presenting participants with single letters or as letters flanked by one of two other letters to the left and right. Often, the aim of such experiments is to investigate limitations of the perceptual system relating to visual acuity, visual angle, or physical size of the stimuli (Hancock, Sawyer & Stafford, 2015) and visual crowding, a phenomenon of neighboring letters seeming to merge perceptually, resulting in misidentification (Bouma, 1970; Bernard et al., 2016; Beier, Bernard & Caster, 2018).

Different models exist to explain the process of word recognition (for example see Davis, 2010; Davis & Bowers, 2004; McClelland & Rumelhart, 1981; Whitney, 2001). In general, it appears that word recognition involves the activation of relatively flexible letter position coding. For example, some models propose that a letter in a specific position (e.g., “o” is in the second position of the word “goat”) will activate the node representing a letter in that specific position as well as other nearby positions (e.g., also the third position and to a weaker degree the fourth position, etc.) (Davis & Bowers, 2004), whereas some other models propose that within-word letter pairs are activated (e.g., the letter pairs “go”, “oa”, “gt” will be activated for the word “goat”) (Grainger & van Heuven, 2004; Snell et al., 2018). This will in turn activate lexical representations of other words with similar letters. Next, whole word representations are mapped onto semantic information in the lexicon (Holcomb & Grainger, 2007).

Word recognition research can use a variety of methods to understand the cognitive processes that subserve word recognition. For example, in lexical decision tasks, participants are presented with real words and either pseudowords or nonwords one at a time and are asked to indicate whether the stimulus in each trial is a word or not. Pseudowords are strings of letters that do not form a word but follow orthographic and phonological rules of the language so they are pronounceable (e.g., “pable”). Nonwords are strings of letters that do not follow orthographic and phonological rules of a language and are unpronounceable (e.g., “pbtlk”). Through the use of single, isolated words and pseudowords/nonwords researchers can probe specific questions that are easier to examine in a more tightly bound context compared to a task using word recognition in a sentence context.

Researchers may also use masked priming tasks where a prime is presented for a short period of time and is masked by either a forward or a backward mask (often a row of hashtags “####”) to ensure the prime isn’t consciously perceived. A target is presented next and they are asked to make a decision about the target, and this is often a lexical decision. Manipulating features of the prime and target allows researchers to investigate the influence of various orthographic and phonological factors on word recognition. These types of experimental tasks can be conducted using behavioral methods where longer reaction times, and potentially lower accuracy, are indicative of more effortful processing. ERPs may also be used to investigate differences to specific ERP components of interest (see Section 5.2.2 for more information).

Methodologies of the letter and word identification may be used to investigate many aspects of orthographic and phonological processing, such as measuring the legibility and readability of a given font, how the text is, itself is laid out on the page, and even how recognizable or familiar the font is to a given reader. Visual features of the background that the text is presented on can also be manipulated.

### 6.1.6. Visual Search Success

Reading is not always a linear, sequential task (e.g., reading through a paragraph in order); readers often have to find a given word or phrase or concept in a text, and while this is reading, it represents a very different task than reading a paragraph from start to finish. Drawing from the cognitive psychology literature, this would be considered a visual search task; that is, looking for a target (for example, a particular word, phrase or even a concept) among many distractors. This question has been the focus of extensive basic research in the study of visual attention (c.f., Treisman and Gelade 1980), and can be broadly thought of as "how does an observer - in our case, a reader - find what they are looking for?"

While the breadth of this literature is outside the scope of this work, Guided Search (Wolfe, Cave, & Franzel, 1989; Wolfe, 2021), which frames our question in terms of the similarities and differences between the target and the distractors, and uses the similarities to guide where the observer attends, is a promising place to start. It is essential to think of search as less "reading" in a more conventional sense, but more of an object identification problem in many ways, and it can be influenced by a range of visual factors in presentation (e.g., font, spacing, density, visual crowding), and by cognitive and linguistic factors. One can imagine, for example, as a reader to find a word in a language they do not read - this would certainly be visual search in a reading context, but quite outside the scope of most reading tasks. More generally, as discussed elsewhere, readers are likely to transition between searching for something specific in a larger text and reading in more depth, and understanding this initial search behavior is key for guiding and helping readers.

### 6.1.7 Pleasure and Preference

Reading for pleasure is a neglected measure of readability, in a literature more likely to focus on speed and accuracy, and indeed we speculate pleasure may be a principal reason for a great part of all reading. Reading for pleasure is documented as a primary reason for purchasing e-readers, as opposed to schools and work environments (Pew Internet Center, 2012), although this may be because users do not find work-related readings straightforward when using e-readers, which do not support easy navigation, annotation, and simultaneous accessibility to multiple documents (Massimi, M., et al., 2013). Leisure reading, conversely, requires a simpler set of functionalities, and with these needs met the digital reading experience must provide reading pleasure (Hancock. Pepe, & Murphy, 2005). Indeed, it is notable that very few evaluations of readability explore this dimension (see Argarwal & Meyer, 2009 for an exception), given that reading has the potential to be an actively and intensely pleasurable act. We suggest this is a future dimension of research, and here fall back to the more modest goal of determining preference.

Font preference is inherently subjective (Miniukovich 2019), and deriving a user's preference is no easy task. There are over 600,000 digital fonts available, and time and attention constraints make the evaluation of even 100 fonts a challenge. O'Donovan et al. identified the struggles graphic designers have when selecting their preferred fonts during real-world tasks. Given designers might have more domain expertise in fonts than the average users, what are more straightforward tasks to derive font preference? A participant’s labelling of preference for various font attributes could be explicit like a Likert scale score from “very much preferred" to “dislike" (Yannakakis & Hallam 2011), or implicit in the case of selecting from two different stimuli in a pairwise comparison. Both approaches have advantages. Prior reading studies have most commonly used Likert scales to determine participant font preference (Banerjee 2011, Bernard et al. 2003, Bhatias et al. 2011, Rello et al. 2016, and Wang et al. 2018). While Likert scales are straightforward, and can be easily averaged across users, when averaging these results they lose their subjective nature (Stevens 1946). Also, the results can be noisy and inconsistent (Negahban et al. 2012) due to a number of factors that are difficult to control for such as visual discomfort per participant (Barkowsky et al. 2018, Mantiuk et al. 2018). Likert scales, with difficult to understand language or representation of a number line (Clark et al. 2018), can pose difficulties for specific populations such as small children (Mellor & Moore 2014) or members of other cultures (Bernard & Gravlee 2014).

A promising alternative to Likert scales are pairwise comparisons (Li et al. 2018), which are more stable because this method is not affected by irrelevant alternatives (Ailon 2008). Human-Computer Interaction researchers have used pairwise comparisons to derive a definitive ranking for a user's preference (Guo et al. 2010, Park et al. 2015, Qian et al. 2015, Yi et al. 2013). For example, Boyarski et al. arranged two physical monitors side-by-side for users to make pairwise comparisons between different fonts (Boyarski 1998). Recently, Wallace et al. designed a digital toggle test to allow users to make pairwise comparisons between different fonts on a single screen (Wallace et al. 2020 chiLBW). allows researchers to definitively rank large numbers of fonts in terms of preference using methods such as Elo Rating or TrueSkill. Pairwise comparisons for a large number of stimuli can take longer given the total number of comparisons a participant must complete. This method can suffer from the transitive property where a participant could prefer font A > B > C > A. Another disadvantage of pairwise comparison is there is currently no accepted hypothesis test available. Recent readability work by Wallace et al. have used a double-elimination tournament to eliminate the transitive property and limit the number of comparisons between 16 different font pairings. This method has an additional benefit where a participant will make comparisons for their more preferred stimuli. There are several other algorithmic efforts to reduce the number of comparisons a participant must complete. These efforts often focus on synthetically completing a pairwise matrix (Kou et al. 2016) or other adaptive approaches (Qian et al 2015).

Future work exploring different methods to derive preference or affinity or font attributes are necessary. While prior work has studied preference between fonts or line-spacing (Rello et al. 2016 makeItBig), as discussed in Section 4.2, there are several additional font attributes that could possibly be explored ranging from character spacing, word spacing, to character width.

## 6.2. Other Readability Methodology Considerations

Readability studies are attended by a number of specific considerations which set them apart from other types of studies. Here, we attempt to capture some of the most common issues that we feel are applicable to reading studies specifically.

### 6.2.1 The Method of Constant Stimuli vs Thresholding

An enduring feature of large-scale readability research is the large individual differences seen between participants (Wallace, et al., 2020). Readability researchers should keep such differences in mind when choosing between two broad methods to measure responses: the method of constant stimuli, or thresholding. The method of constant stimuli dates to the beginnings of experimental psychology and is straightforward (Spearman 1908; Sanford 1888, American Journal of Psychology). The researcher chooses levels of stimulus parameters based on predefined assumptions. Outcome data from such techniques allow for the estimation of psychophysical functions that map the relationship between stimulus levels and performance. However, data collection is limited by the number of trials that can be tolerably collected in a session (the more stimulus levels tested, the more trials required). Stimulus levels must be well chosen for the intended audience; e.g., a text contrast that is reasonably challenging for a younger participant may be too difficult for an older participant.

Researchers may instead choose to employ thresholding or “staircase” procedures. With these methodologies, parameters of the stimulus are adjusted in real-time based on participants’ responses, with the goal of converging on a preselected response accuracy level. Staircasing rules (Levitt 1971; Leek 2001) can be employed to converge on several different accuracy levels. For example, if the experimental task is made more difficult immediately after a participant’s correct response, and made easier by the same amount after an incorrect response, the experiment will eventually converge on a stimulus level representing the participant’s 50% accuracy threshold. A threshold performance value can be determined for every participant without “wasting” trials with parameters that are too difficult or trivially easy. Techniques such as QUEST have updated the general thresholding procedure with more advanced statistical assumptions, allowing for faster convergence (Watson & Pelli 1983). However, if the “step” of the staircase (the amount by which stimulus difficulty is adjusted) is poorly chosen or if the staircase is initialized far from threshold values, it may fail to converge on a good threshold estimate. It can also be more difficult to estimate a full psychometric function from threshold data (Treutwein 1999).

The method of constant stimuli and staircasing are two sides of the same coin. The former holds stimulus parameters constant while measuring changes in performance accuracy; the latter changes stimulus parameters in real-time while holding accuracy constant. Both have their place in the toolkit of legibility research. For an excellent detailed review of such methods, see Klein 2001, Perception & Psychophysics. Legibility researchers continue to use the method of constant stimuli (e.g., Dufau et al 2011; Reimer et al 2014; Sawyer et al 2020 complex backgrounds) as well as thresholding methods (e.g., Roethlein 1912; Sheedy et al 2005; Dobres et al 2016; Sawyer 2020 “Bake Off”).

## 6.2.2 Time on Task, Fatigue, And Vigilance Decrements

We do not read equally well all of the time, and so studies of readability must be sensitive to fluctuations of individual or aggregate ability. People’s alertness levels vary throughout the day, and over the course of a task. Fluctuations in alertness affect cognitive performance and impact higher level cognitive capacities, including perception, memory, and executive functions (Kleitman, 1923). Hence, the ability to concentrate over the course of a study is subject to participants’ alertness levels. A lack of alertness can manifest itself in repeatedly re-reading sentences, troubles with comprehension, and visual fatigue symptoms. Indeed, some tasks are highly demanding and produce their own fatigue, while still other ‘vigilance tasks’ create specific problems for human information processing which grow over time. This “vigilance decrement” is often conflated with fatigue, but is a discrete effect which grows with time on task. Vigilance effects are characterized by simplicity of stimuli, high rates of stimuli evaluation, and low rates of finding whatever “target” is the goal. These types of tasks are often created inadvertently, require hard work, and are intensely stressful for participants, not qualities most scientists are looking to create in their experimentation (Warm, Parasuraman & Matthews, 2008).

In considering time on task, eye trackers can be used to detect participants’ vigilance by tracking blinking rates, for example. Duration and frequency of blinks go up the more tired people generally are. Another way of testing alertness is the use of a psychomotor vigilance test, which tests the ability to sustain attention over time as much as psychomotor speed (Dinges, 1985). We designed and released an open-source test battery for collecting data on people’s varying alertness levels that can be adapted for reading studies (Dingler, 2017). Conducting reading sessions at ‘reasonable hours’ during the day, i.e., avoiding the early morning hours and the so-called ‘post-lunch’ dip, is advisable. Consumption of caffeinated drinks can affect alertness levels as well and should be controlled for (e.g., asking whether participants have had a caffeinated drink in the last hour). Howarth and Costello designed a survey to assess the subjective experience of visual fatigue symptoms, among others (Howarth, 1997). Having participants fill in such a survey before study begins may alert the experimenter to possible confounds through a lack of vigilance. Taking frequent breaks before and between reading sessions is advisable.

### 6.2.3 The Value of Pilot Studies

It is essential to run pilot studies when studying readability. Pilots inform the researcher of positive or negative aspects of their study, and indeed help researchers to understand whether it is worth pursuing the current design. As a researcher, respecting participants means respecting their time. Having participants engage in a study that does not work because it was not piloted is, arguably, ethically problematic. Hence, pilots are beneficial to researchers and future participants alike.

It is essential to spend time and critical thought imagining how individual factors might affect your study, then prove these out in simple pilots. Time spent reading, reading positions, and reading passages can affect the participants across different study environments. How long does it take for participants to learn the interface and become comfortable? At what point in the study do readers naturally slow down or speed up? In a lab study with an eye-tracker and a headrest, participants might become uncomfortable quicker than if they were reading while laying down on their couch at home in a remote study. These factors affect behaviors and are multiplicative with other factors, such as whether a user is moving while reading or facing distractions.

Pacing across a study can also be refined through pilots. Study length of reading studies can vary significantly across participants. Large individual differences in reading speed can cause these differences. One participant could read at 100 words per minute while another reads at 700 words per minute. If the participant has to read 7000 words in a given study, one participant could take 10 minutes to read all the words while another takes 70 minutes. Understanding if a study's reading passages elicit these large differences can drastically affect the number of words a participant reads in a single study. For example, in recent work, readers have to slow down when reading passages normed to a higher reading grade level with more difficult comprehension questions. Understanding how participants slow down their reading speeds to answer comprehension questions will affect the average time they complete a study. Pilot studies can help to understand how many passages a participant needs to read before reading at a comfortable pace. Therefore, each study's training phase can be affected by the reading passages' difficulty and length. Can allow a researcher to adjust failures across these factors and more, resulting in better data and a clearer understanding of the question asked.

Another vital contribution of pilot studies is to help to determine the appropriate compensation provided to participants. As with many studies, monetary compensation is tied to the average time to complete. To provide fair and adequate compensation, a researcher has to make a difficult choice to compensate a participant based upon the average study time across all participants or the time it takes that single participant to complete the study. Compensating slower readers more might provide them an extrinsic motivator to slow down purposely. While extensive literature on compensation exists, in the field of readability there is little precedent for what is “correct”, and so pilots at varying levels of compensation may help researchers to zero in on ideal levels of participant compensation.

When running readability study pilots, it is worth adding additional layers of data collection probing the experience of the reader subjectively. Asking your participants how they felt, where they were confused, whether compensation was sufficient, or whether instructions were clear can be vital to understanding the root cause of difficulties with data. We suggest asking these questions after the post survey, and if possible engaging in a free-form conversation with some or all pilot participants. We also recommend engaging in your own research, in full. Researchers have many assumptions about the arc of their own research which can be challenged, and corrected, by putting themselves through all of the steps that their participants will experience.

# 7. Research Design and Data Analysis

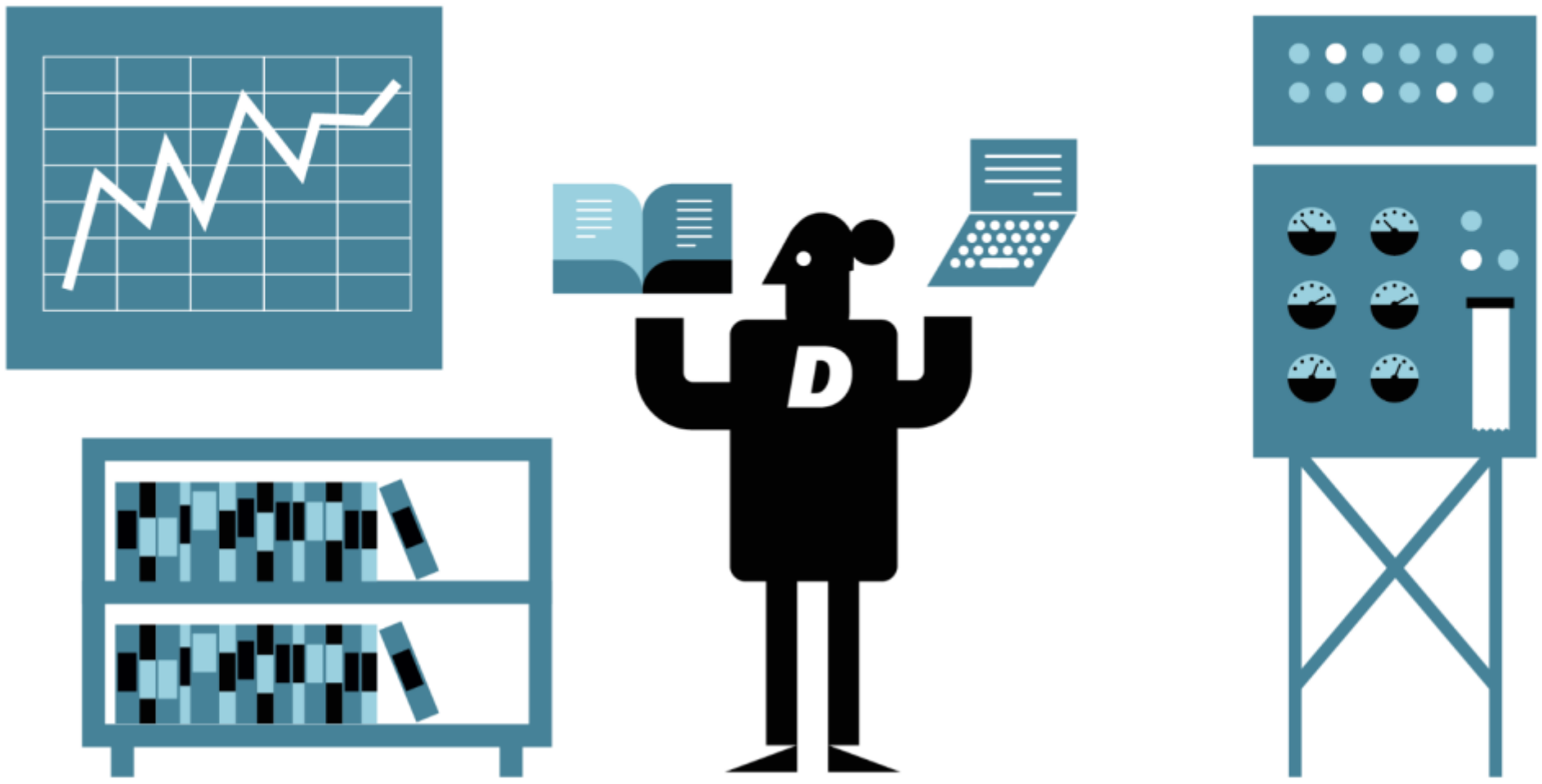

**Figure 15.** Readability is highly multidisciplinary, and so design and analysis approaches vary widely.

Not everyone who studies readability comes from the same scientific background, or indeed comes from a scientific background at all. Therefore, the present section provides an aspiring readability researcher an overview of common research design, data quality management, as well as statistical and machine learning modeling approaches. These examples have been gathered from existing readability research, but as the area is growing and multidisciplinary, the section is not intended to be an exhaustive list of the approaches which might be useful for readability researchers (Fig. 15). We hope that the pooled experiences of the authorship provide useful starting points, extensions, and counterpoints for individuals at a variety of levels of experience, and a variety of fields.

## 7.1 Common Research Design

Methods for readability studies are diverse, and the exhaustive list may be as broad as all possible populations in all possible contexts. Therefore, we here briefly address three common methods which cover a useful range of reading contexts.

### 7.1.1 Readability at-a-glance

In many applications, such as reading street signs, directions, or text on the face of a watch, high speed and comprehension is essential for safety and efficiency. This regime is termed "glanceable reading, as reading occurs in extremely short periods of time, often less than one second. Performance in reading single words was studied by Dobres et al. using a "staircasing" method to evaluate fonts and how fast single words can be read in them (Sawyer et al., 2017; Dobres et al. 2016). This method presented single words or pseudo-words (e.g. "thigma"), similar to the word recognition task employed by Meyer and Schvaneveldt (1971). Such exploration of at-a-glance reading reveals valuable insights about optimal design for the kinds of reading that are done at speed.

### 7.1.2 Readability in Opportunistic Interludes

Reading of text that is in multiple sentences or longer falls in the regime of interlude reading or long-form reading. Reading relatively short blocks of text often occurs in interludes. These are slices of time between other activities or concurrent with other activities in a multitasking frame (e.g. Reeves et al. 2020). For example, while waiting in a queue, people will commonly consume news or social information for a few minutes. Methods for studying speed and comprehension in this regime differ from the psychophysical approach used to investigate glanceable reading. To study reading of short passages of text, Wallace et al. (Towards Readability Individuation) presented participants with passages of text in varying fonts, and measured how long it took for participants to read the passages. Each passage set was followed by a set of comprehension questions, to encourage participants to read with adequate comprehension, and providing a measure that could be used to disqualify data that did not meet a standard of adequate attention to the reading task. They found that participants' fastest font could allow reading up to 32% faster than an individual's slowest font, with relatively stable and high comprehension, between 86 and 94%. While the passages used by Wallace et al. were relatively simple, a similar method of timing reading of short passages can be used to investigate reading speed and comprehension of somewhat longer or more complex reading passages just as easily.

### 7.1.3 Readability in Long Documents

Longer texts, such as academic articles, legal briefs, or novels can be considered long-form reading, where reading is the focal task and any distraction from it is a secondary task. The methods employed by Wallace et al. (2020) are likely to be only partly applicable, as one often does not read an entire book in an uninterrupted epoch, as one can read a single paragraph or page without stopping. More sophisticated measures to assess the impact of type features on reading speed and comprehension in long form reading are needed, following research that has been ongoing for quite a long time in the reading science and cognition spaces.

## 7.2. Data Quality Management

Readability studies rely heavily on data quality. Because many reported effects in readability are small to medium in effect size, it is necessary to repeat many trials within individuals, or collect very large groups of individuals. Both situations provide plenty of opportunities for data quality issues. Not all participants perform their task with the same level of dedication, and there are high levels of individual differences in reading ability and strategy, both of which can result in data anomalies, or outliers. Defining and detecting outliers is something of an art, and must be tailored depending upon the study design and population. In studies which involve many trials within an individual, manipulation checks for effects of time on task or training effects must be undertaken. In studies with a large number of participants, especially those conducted on crowdsourcing platforms, care must be taken to screen for participant dishonesty (Peer et al. 2017), uncontrolled settings (Schneegass and Draxler 2021). In both cases, and in all reading and readability studies, individuals at both extremes of reading abilities skew the analysis (Carver 1990). In general, data quality issues can be mitigated through careful planning, piloting, and in studies with significant data collection time, ongoing data quality assurance.

A common statistical approach to handling outliers is to assume normal distribution on the data and isolate points that fall 3+ standard deviations from the mean (Stevens 2012). These points can then either be filtered out or reported on separately, depending on the experiment. Researchers can also leverage anomaly detection methods during data pre-processing. For instance, in the case of reading speed data, outlier removal can be done based on domain knowledge of expected reading speed distributions. Typical reading speeds for participants over the age of 18 range from 138 to 600 wpm (Carver 1990) with an average speed for native English speakers at 240 words per minute. Participants whose speed falls outside of this range might be distracted or disengaged from the material, and may be removed from the analysis. Of course, it is also important to pilot and understand what “good” looks like in the particular context of the study at hand.

Variability within a particular participant’s data poses a significant challenge for analysis, and should create concern for similar patterns across participants in the entire study. When a similar task is repeated by an individual multiple times, the random error associated with the repeated measurement of independent performance factors, such as attention (Raichle et al., 2001; Buckner et al., 2008; Christoff et al., 2009; Killingsworth and Gilbert, 2010), can attenuate the association between independent and dependent variables and result in poor statistical inference (Barnett et al., 2005), a bias known as regression dilution (Hutcheon et al., 2010; Berglund, 2012). In general, unusually high intra-participant variability may also be a sign of problems with experimental design, and common confounds such as unmet training requirements, excessive time on task, uneven population reading ability, and technical failures should be investigated, among others.

## 7.3. Exploration and Visualization

Readability data is best initially analysed through Exploratory Data Analysis (EDA), which can help determine data quality, assist in numerical analysis, or build hypotheses for further investigation. Exploration must take into account the nature of readability data, and must respect any *a priori* plan for analysis. While inspection of raw readability data in a tabular format can actually be revealing, we have found a few specific visualizations helpful for revealing interesting patterns. Because readability data is often collected over time, across environments, and between devices, we see opportunities to use spatio-temporal illustrations to explain the complex emerging patterns, heatmaps to visualize one-time movement patterns shared across users, or sankey diagrams to incorporate higher levels of complexity in participants' shared journeys across different stages of the reading process. We believe many other novel visualizations can also help represent eye and mouse movement data on the reading interface, where the areas with more attention can be highlighted by warmer colors (Blignaut, 2010; Burch et al., 2019). When a common reference frame such as the location at the start of the text is defined, proximity-based visualization can reduce the spatial dimension on a regular 2D map, integrating the temporal dimension for easy comparison across users (Crnovrsanin et al., 2009).

## 7.4. Statistical Modeling

So long as the tools are used appropriately relative to the data collected, we do not yet see any strong place for the dominance of any statistical approach.The data analyses needed for readability experiments examining the effect of visual manipulations on outcomes such as speed, comprehension, and preference are similar those used throughout the social sciences. The standard practice for statistical analysis is to start with numerical and graphical techniques for estimating the distribution of the data and determining the best mechanism accordingly. A simple Kolmogorov- Smirnoff test can determine whether readability scores such as reading speed and comprehension are normally distributed. Parametric tests, in the case of normality, and non-parametric tests, otherwise, are often used in the readability research studies (Soleimani, 2009; Soleimani et al. 2012), and indeed the non-normality of distributions of many metrics may not be cause for concern, so much as cause for use of the appropriate tools.

There is no prescriptive statistical tool for readability research, which is commonly analyzed with multiple generalizations of the general linear model (GLM). Many studies rely upon multiple analysis of variance (MANOVA) to isolate the impact of independent variables (IVs) upon multiple dependent variables (DVs), often including both reading speed and comprehension (Wallace et al. 2020; Nam et al. 2020; Rello, Pielot, and Marcos 2016; Sawyer et al. 2020; Gao et al. 2019). It's not uncommon to augment these larger analyses with smaller "manipulation checks" which rely upon t-tests or simple analysis of variance (ANOVA) to test out assumptions.

In determining how manipulations affect populations across continuous variables, it is certainly appropriate to use regression analysis.  Indeed, in usability studies where a simple question of "A or B" is of interest, and where multiple DVs are not used, a simple t-test can suffice. Readability, as an inherently multidisciplinary area of inquiry, should ultimately be modelled using the tools most familiar to the researcher investigating.

## 7.5. Machine Learning

For research that aims to assist participants in improving readability, it can be useful to evaluate the performance of statistical and machine learning (ML) models that can predict reading outcomes. Consider one question in the literature: given a font style, can a participant's reading speed be predicted (Cai & Wallace, 2021)? Here, regression models which predict the relationship between input and output variables might be used to predict participants' reading level from their reading speed. This regression question would be valid in statistical and ML approaches alike, and indeed the outcomes of these two approaches might be notably similar. Classification ML allows the prediction of a label for a given set of input variables, and so in the context of readability might be useful for predicting the "bin" into which such an input set might fall. A simple binary classification might detect whether a participant is skimming, or reading deeply, given their reading speed. Similarly, using ranking ML, a given set of input can predict an ordered set of labels. . In more sophisticated learning models, ranking ML can predict the relative ordering of labels by either comparing pairs of inputs at a time, or by comparing the entire set of labels associated to our criterion.(Liu 2007). Consider ranking the fonts for each person in a way that the most readable font for a particular person is ranked first, and the least readable is ranked last. Clustering ML groups similar items together, perhaps providing groups of similar readers and identifying populations in need. A full survey of traditional approaches can be found in (Rui Xu and Wunsch 2005) with authors often using the classical approach of K-means. More recently clustering research has focused on a subfield known as metric learning that learns a feature representation where neighboring items are closer together in feature space. Of course, as with statistical tools, ML approaches are best used together to achieve complex goals of prediction.

ML tools in the family of so-called Deep Learning approaches, multi-layer and often convolutional ML which advance the state-of-the-art for each approach named above, have special considerations (LeCun, 2015). Data hungry, building models using these approaches is challenging for small and medium datasets. When properly attached to truly big data, these methods do allow very large parameter models to optimize a loss function, thus maximizing prediction accuracy, they are challenging in terms of transparency.  Indeed, what these models give in prediction they take away in terms of understanding the causal reason for their explanation, and specifically in terms of understanding which features are important (Samek, 2017). While there is much work to bypass these shortcomings, for the moment more traditional ML approaches may be a better path for researchers interested in understanding why their models function, and building their ML approaches out of modestly-sized datasets.

## 7.6. Complementary Expertise

One common experience that unites the authorship of the present work is the great benefit to multidisciplinary working groups investigating readability. Readability is a complex topic, touching upon engineering, physiology, neuroscience, psychology, social science, education, and still more fields. Experts come with their own viewpoint, their own tools, and their own recipes for research design and data analysis. There is a lot of give-and-take in the most successful readability projects, which to our mind is the point: the most valuable revelations in this topic area are certainly at intersections between established fields. As such we encourage as follows: explore together. If research is not your first calling, find a supportive researcher. A graduate degree which includes research goes far beyond the scope of this section. If machine learning is not your first calling, strike up conversations with those with a talent for this rapidly emerging toolset. If design and typography are foreign to you, have conversations with those that are fluent in the building blocks of readability. All this is not to say that a determined individual cannot design good readability research and adequately analyze the resulting data alone. We simply suggest that this journey may be more rewarding for diverse, like-minded partners.

# 8. A Call for Further Research

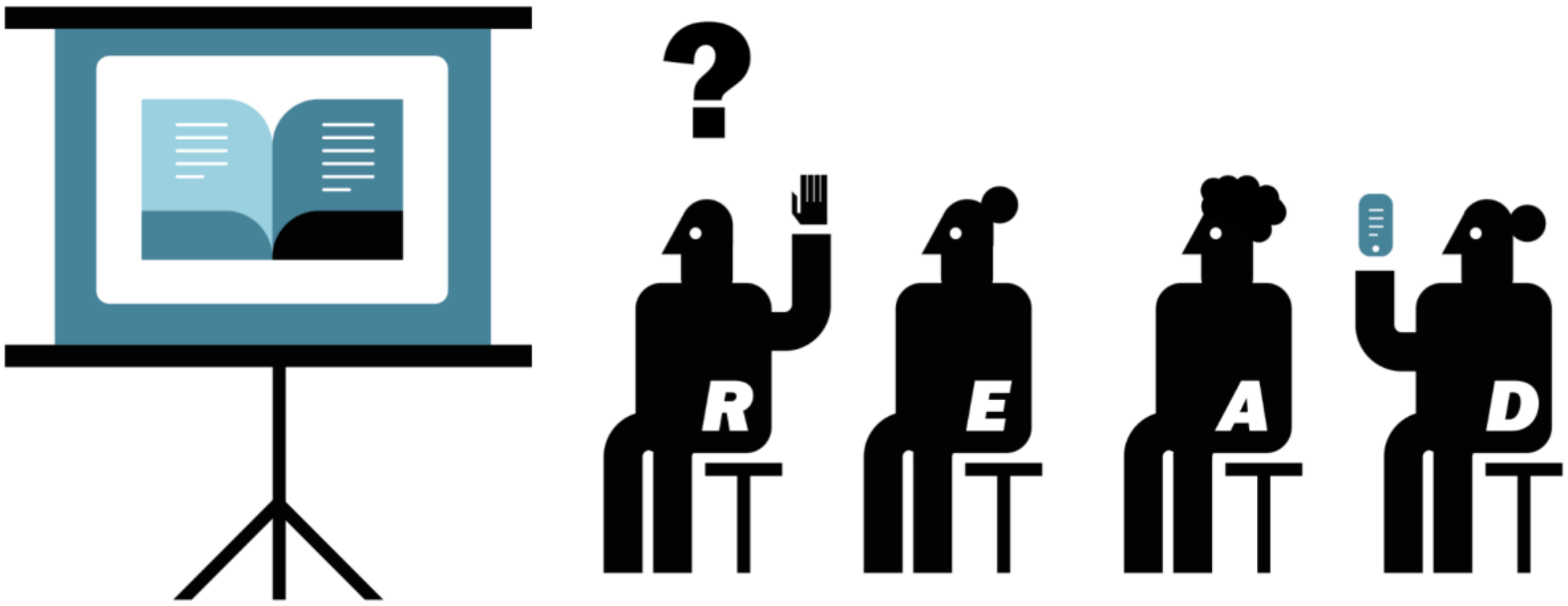

**Figure 16.** The growing readability research community is still early in building understanding.

Digital devices and the way they are connected are rapidly changing the availability and access to information. Now more than ever, the ability to read - and to do so efficiently - has a direct and dramatic impact on education, health, and career outcomes. At the same time, digital devices provide an opportunity to create new personalized reading environments to build capacity for all readers, including proficient and non-proficient, and even those who do not think of themselves as readers at all (Fig. 16).

**Rather than one-format-fits-all, there is an opportunity to individuate both for the reader and for the reading context.** Prior reading studies have contributed significantly to our understanding of how typographical variables affect readability. However, prior work has focused on the idea that big change, such as font, can benefit all readers. We, however, suggest a shift: modern reading research must evaluate how small changes to text format on an individual basis can create significant outcomes for the reader and that these variations of text may be different, depending on the content, device and reading context. As devices, applications, design trends, and typography have evolved, we can revisit prior research to adjust or expand upon previous findings.

**Multidisciplinary shared research is required.** Readability is a complex field of research that requires a multidisciplinary approach. In this paper, we provided a taste of the elements that must come together to form a readability study, including: the preparation of reading materials and typographic decisions; the selection of study participants and considerations for human subjects research; the hardware devices and software platforms on which the studies can take place, together with instrumentation for evaluation (e.g., eye tracking and neuroimaging); the experimental methodologies that can provide a systematic evaluation of reading behaviors across study conditions; and the interpretation of the reading data to build both inference and predictive models.

Readability studies can range in complexity from simple timed studies to sensor-heavy scientific investigations, from small scale laboratory studies to large online studies with hundreds or thousands of participants; and such studies can be focused on individuals, specific subpopulations or mainstream populations, reading on desktop, mobile, or wearable devices. Therefore, an understanding of perceptual science, human factors, reading subject matter expertise, design, neuroimaging, statistics, software engineering, sensors and systems as well as machine learning must come together to use sound methodologies across disciplines to craft meaningful experiments and experimental platforms as well as correctly interpret results.

**There is a need for research, investment, and standardization.** The authors call for further research and study, and investment by education, industry, government, and policymakers. Scientists and practitioners from industry and academia, as well as creatives and type designers can come together to make better readability a reality for all. The scope of readability research is vast, and the methods are varied. We wrote this paper to map out the best practices and methodologies available to readability researchers to both inform and hopefully to standardize both practice and data collection.

**Publicly available data and tools are required for reproducible readability research.** We urge communities of researchers, engineers, and designers to release reading content, typography, experimental designs, software platforms, analysis tools, and computational models. This will allow other groups to benefit from subject matter expertise, to run more controlled and reproducible studies, and to compare results across populations, context, and settings by virtue of a common set of tools. As we develop recommendations of formats, mapping individual characteristics to readability features, the benefits should be made available to all readers by all relevant technical partners. Together, let us engineer better reading for everybody.

**Join the readability research community.** The authors invite you to collaborate:

- Make use of the free Virtual Readability Lab toolset for testing reading populations. Contact Dr. Ben D. Sawyer for more information.
- Contact the team at Readability Matters to have your research highlighted on the Readability Matters Research page.
- Follow the work of The Readability Consortium.

# Author Contributions

This paper was generated collaboratively over the course of a series of online workshops, the results of which were extensively edited by Dr. Zoya Bylinskii, Dr. Ben Sawyer, and Dr.Benjamin Wolfe. Original illustrations by Bernard Kerr.

# Acknowledgements

We acknowledge the following individuals for their valuable contributions to this manuscript: Sarah Barrientos, Jose Echevarria, Xander Koo, Max Rose, Xi Wang.